\documentclass[aps,prl,twocolumn,groupedaddress]{revtex4-2}
\usepackage{xcolor}
\usepackage{amsmath}
\usepackage{dsfont}

\newcommand{\average}[1]{\langle #1 \rangle}
\newcommand{\tr}{\text{tr}}
\newcommand{\Monedim}{\mathfrak{m}_\mathcal{R}}
\usepackage{graphicx}
\usepackage{subcaption} % to use subfigure environment (pics side by side)
\usepackage{xcolor}
\usepackage{amsmath,amssymb,amsthm}
\usepackage{bm}
\usepackage{float}
\usepackage[format=plain,labelfont={bf,it},textfont=it]{caption}
\usepackage{multirow}
\usepackage{slashed} % for Feynman slashed notation
\usepackage{mathtools} % for subequation numbering
\usepackage{tabularx}
\usepackage{bbold}
\usepackage[shortlabels]{enumitem}
\usepackage{tikz}
\usetikzlibrary{positioning,arrows,calc}
\usetikzlibrary{positioning,arrows,calc}
\usetikzlibrary{decorations.pathmorphing} 
\usepackage{dsfont} % for displaying the blackboard 1 properly
\newcommand{\DA}[1]{\textbf{\textcolor{orange}{[DA: #1]}}}
\newcommand{\CM}[1]{\textbf{\textcolor{blue}{[CM: #1]}}}

\newcommand{\vev}[1]{\left\langle #1 \right\rangle}

\newcommand\x{.65}

\newcommand\vertexsize{.06}

\definecolor{DarkBlueGrey}{RGB}{76,94,107}
\definecolor{MediumBlueGrey}{RGB}{110,135,153}
\definecolor{LightBlueGrey}{RGB}{134,163,184}
\definecolor{VeryLightBlueGrey}{RGB}{242,249,255}
\definecolor{WCOrange}{RGB}{242,146,29}
\definecolor{VeryLightOrange}{RGB}{255,245,233}
\definecolor{SCRed}{RGB}{179,48,48}
\definecolor{VeryLightRed}{RGB}{255,239,239}
\definecolor{VertexColor}{RGB}{242,146,29}
\definecolor{GluonColor}{RGB}{255,172,172}
\definecolor{SEColor}{RGB}{134,163,184}

\tikzset{
corner/.style={line width=1pt,dashed,draw=black,dash pattern=on 6pt off 4pt},
scalar/.style={line width=1pt,draw=black},
gluon/.style={line width=1pt,decorate, draw=GluonColor,
    decoration={complete sines,aspect=0,amplitude=1.25mm,segment length=1.5mm,start up,end up}},
gluontwo/.style={
    line width=1pt,
    draw=GluonColor,
    decorate, 
    decoration={snake, amplitude=.7mm, segment length=1.5mm}
  },
ghost/.style={line width=1pt,loosely dotted,draw=black},
wilson/.style={line width=2pt,draw=black},
 }

\tikzset{
    tracecircle/.style={
        draw=DarkBlueGrey,
        line width=1.5pt
    },
    contraction/.style={
        draw=SCRed,
        line width=1.3pt,
        line cap=round
    },
    contractionover/.style={
        preaction={draw=white,line width=3.2pt},
        draw=SCRed,
        line width=1.3pt,
        line cap=round
    }
}

\NewDocumentCommand\semiloop{O{black}mmmO{}O{above}}
{%
\draw[#1] let \p1 = ($(#3)-(#2)$) in (#3) arc (#4:({#4+180}):({0.5*veclen(\x1,\y1)})node[midway, #6] {#5};)
}

\newcommand{\DefectSelfEnergy}{\begin{tikzpicture}[baseline={([yshift=-2ex]current bounding box.center)}]
\draw[wilson] (0,0) -- (1.5,0);
\draw[fill=white] (.25,0) circle (.075);
\draw[fill=white] (1.25,0) circle (.075);
\path[clip] (0,0) rectangle (1.5,1);
\draw[scalar] (.75,0) circle (.5);
\draw[wilson] (0,0) -- (1.5,0);
\draw[fill=white] (.25,0) circle (.075);
\draw[fill=white] (1.25,0) circle (.075);
\draw[fill=SEColor,SEColor] (0.75,.5) circle (.1);
\end{tikzpicture}}

\newcommand{\DefectSSSSSelfEnergyOne}{\begin{tikzpicture}[baseline={([yshift=-2ex]current bounding box.center)}]
\draw[wilson] (0,0) -- (2,0);
\draw[fill=white] (.25,0) circle (.075);
\draw[fill=white] (.75,0) circle (.075);
\draw[fill=white] (1.25,0) circle (.075);
\draw[fill=white] (1.75,0) circle (.075);
\path[clip] (0,0) rectangle (2,1);
\draw[scalar] (.75,0) circle (.5);
\draw[scalar] (1.25,0) circle (.5);
\draw[wilson] (0,0) -- (2,0);
\draw[fill=white] (.25,0) circle (.075);
\draw[fill=white] (.75,0) circle (.075);
\draw[fill=white] (1.25,0) circle (.075);
\draw[fill=white] (1.75,0) circle (.075);
\draw[fill=SEColor,SEColor] (.5,.4) circle (.1);
\end{tikzpicture}}

\newcommand{\DefectSSSSSelfEnergyTwo}{\begin{tikzpicture}[baseline={([yshift=-2ex]current bounding box.center)}]
\draw[wilson] (0,0) -- (2,0);
\draw[fill=white] (.25,0) circle (.075);
\draw[fill=white] (.75,0) circle (.075);
\draw[fill=white] (1.25,0) circle (.075);
\draw[fill=white] (1.75,0) circle (.075);
\path[clip] (0,0) rectangle (2,1);
\draw[scalar] (.75,0) circle (.5);
\draw[scalar] (1.25,0) circle (.5);
\draw[wilson] (0,0) -- (2,0);
\draw[fill=white] (.25,0) circle (.075);
\draw[fill=white] (.75,0) circle (.075);
\draw[fill=white] (1.25,0) circle (.075);
\draw[fill=white] (1.75,0) circle (.075);
\draw[fill=SEColor,SEColor] (1.5,.4) circle (.1);
\end{tikzpicture}}

\newcommand{\DefectSSSSX}{\begin{tikzpicture}[baseline={([yshift=-2ex]current bounding box.center)}]
\draw[wilson] (0,0) -- (2,0);
\draw[fill=white] (.25,0) circle (.075);
\draw[fill=white] (.75,0) circle (.075);
\draw[fill=white] (1.25,0) circle (.075);
\draw[fill=white] (1.75,0) circle (.075);
\draw[fill=VertexColor,VertexColor] (1,.43) circle (\vertexsize);
\path[clip] (0,0) rectangle (2,1);
\draw[scalar] (.75,0) circle (.5);
\draw[scalar] (1.25,0) circle (.5);
\draw[wilson] (0,0) -- (2,0);
\draw[fill=white] (.25,0) circle (.075);
\draw[fill=white] (.75,0) circle (.075);
\draw[fill=white] (1.25,0) circle (.075);
\draw[fill=white] (1.75,0) circle (.075);
\draw[fill=VertexColor,VertexColor] (1,.43) circle (\vertexsize);
\end{tikzpicture}}

\newcommand{\DefectSSSSH}{\begin{tikzpicture}[baseline={([yshift=-2ex]current bounding box.center)}]
\draw[wilson] (0,0) -- (2,0);
\draw[fill=white] (.25,0) circle (.075);
\draw[fill=white] (.75,0) circle (.075);
\draw[fill=white] (1.25,0) circle (.075);
\draw[fill=white] (1.75,0) circle (.075);
\draw[fill=VertexColor,VertexColor] (.4,.36) circle (\vertexsize);
\draw[fill=VertexColor,VertexColor] (1.6,.36) circle (\vertexsize);
\path[clip] (0,0) rectangle (2,1);
\draw[scalar] (.75,0) circle (.5);
\draw[scalar] (1.25,0) circle (.5);
\draw[wilson] (0,0) -- (2,0);
\draw[fill=white] (.25,0) circle (.075);
\draw[fill=white] (.75,0) circle (.075);
\draw[fill=white] (1.25,0) circle (.075);
\draw[fill=white] (1.75,0) circle (.075);
\draw[fill=VertexColor,VertexColor] (.4,.36) circle (\vertexsize);
\draw[fill=VertexColor,VertexColor] (1.6,.36) circle (\vertexsize);
\path[clip] (0,.36) rectangle (2,1);
\draw[gluontwo] (1,.15) circle (.65);
\draw[fill=VertexColor,VertexColor] (.4,.36) circle (\vertexsize);
\draw[fill=VertexColor,VertexColor] (1.6,.36) circle (\vertexsize);
\end{tikzpicture}}

\newcommand{\DefectY}{\begin{tikzpicture}[baseline={([yshift=-2ex]current bounding box.center)}]
\draw[wilson] (-.2,0) -- (1.75,0);
\draw[gluontwo] (.4,.36) -- (.1,.66);
\draw[fill=white] (.25,0) circle (.075);
\draw[fill=white] (1.25,0) circle (.075);
\draw[fill=VertexColor,VertexColor] (.4,.36) circle (\vertexsize);
\draw[fill=VertexColor,VertexColor] (0,0) circle (\vertexsize);
\draw[fill=VertexColor,VertexColor] (.75,0) circle (\vertexsize);
\draw[fill=VertexColor,VertexColor] (1.5,0) circle (\vertexsize);
\path[clip] (0,0) rectangle (1.75,1);
\draw[scalar] (.75,0) circle (.5);
\draw[wilson] (0,0) -- (1.75,0);
\draw[fill=white] (.25,0) circle (.075);
\draw[fill=white] (1.25,0) circle (.075);
\draw[fill=VertexColor,VertexColor] (.4,.36) circle (\vertexsize);
\draw[fill=VertexColor,VertexColor] (0,0) circle (\vertexsize);
\draw[fill=VertexColor,VertexColor] (.75,0) circle (\vertexsize);
\draw[fill=VertexColor,VertexColor] (1.5,0) circle (\vertexsize);
\end{tikzpicture}}

\newcommand{\DefectSSSSYOne}{\begin{tikzpicture}[baseline={([yshift=-2ex]current bounding box.center)}]
\draw[wilson] (-.2,0) -- (2.2,0);
\draw[gluontwo] (.4,.36) -- (.1,.66);
\draw[fill=white] (.25,0) circle (.075);
\draw[fill=white] (.75,0) circle (.075);
\draw[fill=white] (1.25,0) circle (.075);
\draw[fill=white] (1.75,0) circle (.075);
\draw[fill=VertexColor,VertexColor] (.4,.36) circle (\vertexsize);
\draw[fill=VertexColor,VertexColor] (0,0) circle (\vertexsize);
\draw[fill=VertexColor,VertexColor] (.5,0) circle (\vertexsize);
\draw[fill=VertexColor,VertexColor] (1,0) circle (\vertexsize);
\draw[fill=VertexColor,VertexColor] (1.5,0) circle (\vertexsize);
\draw[fill=VertexColor,VertexColor] (2,0) circle (\vertexsize);
\path[clip] (0,0) rectangle (2,1);
\draw[scalar] (.75,0) circle (.5);
\draw[scalar] (1.25,0) circle (.5);
\draw[wilson] (0,0) -- (2,0);
\draw[fill=white] (.25,0) circle (.075);
\draw[fill=white] (.75,0) circle (.075);
\draw[fill=white] (1.25,0) circle (.075);
\draw[fill=white] (1.75,0) circle (.075);
\draw[fill=VertexColor,VertexColor] (.4,.36) circle (\vertexsize);
\draw[fill=VertexColor,VertexColor] (0,0) circle (\vertexsize);
\draw[fill=VertexColor,VertexColor] (.5,0) circle (\vertexsize);
\draw[fill=VertexColor,VertexColor] (1,0) circle (\vertexsize);
\draw[fill=VertexColor,VertexColor] (1.5,0) circle (\vertexsize);
\draw[fill=VertexColor,VertexColor] (2,0) circle (\vertexsize);
\end{tikzpicture}}

\newcommand{\DefectSSSSYTwo}{\begin{tikzpicture}[baseline={([yshift=-2ex]current bounding box.center)}]
\draw[wilson] (-.2,0) -- (2.2,0);
\draw[gluontwo] (1.6,.36) -- (1.9,.66);
\draw[fill=white] (.25,0) circle (.075);
\draw[fill=white] (.75,0) circle (.075);
\draw[fill=white] (1.25,0) circle (.075);
\draw[fill=white] (1.75,0) circle (.075);
\draw[fill=VertexColor,VertexColor] (1.6,.36) circle (\vertexsize);
\draw[fill=VertexColor,VertexColor] (0,0) circle (\vertexsize);
\draw[fill=VertexColor,VertexColor] (.5,0) circle (\vertexsize);
\draw[fill=VertexColor,VertexColor] (1,0) circle (\vertexsize);
\draw[fill=VertexColor,VertexColor] (1.5,0) circle (\vertexsize);
\draw[fill=VertexColor,VertexColor] (2,0) circle (\vertexsize);
\path[clip] (0,0) rectangle (2,1);
\draw[scalar] (.75,0) circle (.5);
\draw[scalar] (1.25,0) circle (.5);
\draw[wilson] (0,0) -- (2,0);
\draw[fill=white] (.25,0) circle (.075);
\draw[fill=white] (.75,0) circle (.075);
\draw[fill=white] (1.25,0) circle (.075);
\draw[fill=white] (1.75,0) circle (.075);
\draw[fill=VertexColor,VertexColor] (1.6,.36) circle (\vertexsize);
\draw[fill=VertexColor,VertexColor] (0,0) circle (\vertexsize);
\draw[fill=VertexColor,VertexColor] (.5,0) circle (\vertexsize);
\draw[fill=VertexColor,VertexColor] (1,0) circle (\vertexsize);
\draw[fill=VertexColor,VertexColor] (1.5,0) circle (\vertexsize);
\draw[fill=VertexColor,VertexColor] (2,0) circle (\vertexsize);
\end{tikzpicture}}

\newcommand{\TraceGij}{%
\begin{tikzpicture}[
    baseline={([yshift=-.6ex]current bounding box.center)},
    scale=.72
]
    \coordinate (i) at (180:.65);
    \coordinate (j) at (0:.65);

    \draw[contraction] (i) -- (j);

    \draw[tracecircle] (0,0) circle (.65);

    \fill[DarkBlueGrey] (i) circle (.055);
    \fill[DarkBlueGrey] (j) circle (.055);

    \node[left=2pt]  at (i) {$i$};
    \node[right=2pt] at (j) {$j$};
\end{tikzpicture}%
}

\newcommand{\TraceBijkl}{%
\begin{tikzpicture}[
    baseline={([yshift=-.6ex]current bounding box.center)},
    scale=.72
]
    \coordinate (i) at (135:.65);
    \coordinate (j) at (45:.65);
    \coordinate (k) at (-45:.65);
    \coordinate (l) at (-135:.65);

    \draw[contraction]     (i) -- (k);
    \draw[contractionover] (j) -- (l);

    \draw[tracecircle] (0,0) circle (.65);

    \fill[DarkBlueGrey] (i) circle (.055);
    \fill[DarkBlueGrey] (j) circle (.055);
    \fill[DarkBlueGrey] (k) circle (.055);
    \fill[DarkBlueGrey] (l) circle (.055);

    \node[above left=1pt]  at (i) {$i$};
    \node[above right=1pt] at (j) {$j$};
    \node[below right=1pt] at (k) {$k$};
    \node[below left=1pt]  at (l) {$\ell$};
\end{tikzpicture}%
}

\newcommand{\TraceXijkl}{%
\begin{tikzpicture}[
    baseline={([yshift=-.6ex]current bounding box.center)},
    scale=.72
]
    \coordinate (i) at (135:.65);
    \coordinate (j) at (45:.65);
    \coordinate (k) at (-45:.65);
    \coordinate (l) at (-135:.65);
    \coordinate (v) at (0,0);

    \draw[contraction] (i) -- (v);
    \draw[contraction] (j) -- (v);
    \draw[contraction] (k) -- (v);
    \draw[contraction] (l) -- (v);

    \draw[tracecircle] (0,0) circle (.65);

    \fill[DarkBlueGrey] (i) circle (.055);
    \fill[DarkBlueGrey] (j) circle (.055);
    \fill[DarkBlueGrey] (k) circle (.055);
    \fill[DarkBlueGrey] (l) circle (.055);

    \fill[SEColor] (v) circle (.085);

    \node[above left=1pt]  at (i) {$i$};
    \node[above right=1pt] at (j) {$j$};
    \node[below right=1pt] at (k) {$k$};
    \node[below left=1pt]  at (l) {$\ell$};
\end{tikzpicture}%
}
\newcommand{\scaleddiagram}[1]{\scalebox{0.8}{#1}}

\usepackage[sectionbib]{bibunits}
\defaultbibliographystyle{apsrev4-1}
\defaultbibliography{\jobname Notes,letter}

\begin{document}

%\preprint{}

\title{New Exotic Operators in the Spectrum \\ of  Wilson Lines in General Representations}


\author{Daniele Artico}
\author{Carlo Meneghelli}
%\email[]{carlo.meneghelli@unipr.it}
\author{Michele Savi}
\author{Rudolfs Treilis}
%\thanks{}
\affiliation{Dipartimento di Fisica, Universit\`a di Parma, 
and INFN, Gruppo di Parma}

%%% E-mail addresses


\date{\today}

\begin{abstract}
Wilson lines are fundamental probes of gauge theories. We show that, in sufficiently rich representations, they support a large new class of operator insertions. For half-BPS lines in $\mathcal{N}=4$ SYM, some have the quantum numbers of the displacement supermultiplet
and define natural defect deformations. 
By relating their beta-functions to chiral ring OPE coefficients, we show that these deformations are marginally irrelevant, meaning the half-BPS line is IR stable for all representations.
We connect these deformations to multidimensional RG flows and find evidence for  new conformal line defects.

\end{abstract}

%\keywords{}

\maketitle

\begin{bibunit}
\section{Introduction \label{sec:intro}}
Line operators play a prominent role in the study of gauge theories: originally introduced to serve as an order parameter for confinement \cite{PhysRevD.10.2445}, it was soon after realised that, due to their gauge invariance, they act as natural variables for the formulation of these theories \cite{Makeenko:1979pb,Polyakov:1980ca}. Ever since then the Wilson-line operator has retained a position of high importance in the study of gauge interactions. Given a connection $\mathcal{A}$ on a principal $G$-bundle, a representation $\mathcal{R}\in\text{Rep}(G)$, and a contour $\mathcal{C}$, we may define the holonomy around the contour $\mathcal{C}$ as
\begin{equation}
    \mathbb{W}_\mathcal{R}[\mathcal{A},\mathcal{C}] = \text{tr}_\mathcal{R}\mathds{P}\exp\big( i\oint_{\mathcal{C}}\mathcal{A}\,ds\big)\,.
\end{equation}
Since gauge transformations are redundancies in the description of the theory, all well-defined observables need to be gauge invariant. The constraint on local bulk operators is then easily formulated as $\mathcal{O}_{bulk}[\mathcal{A}^g,\Phi^g] = \mathcal{O}_{bulk}[\mathcal{A},\Phi]$, where $\Phi$ are all the matter fields in the theory and $g(x)$ specifies the gauge transformation. In contrast, on the line this condition is relaxed and a failure of a local operator to be invariant can be compensated by the transformation of the line itself. The observables on a line are naturally $\text{End}(\mathcal{R})$-valued, i.e. $\dim(\mathcal{R})\times\dim(\mathcal{R})$ matrices, and the condition for gauge invariance is
\begin{equation}\label{eqn_gauge_line_general}
    \mathcal{O}_{line}[\mathcal{A}^g,\Phi^g] = \rho_{\mathcal{R}}(g)\mathcal{O}_{line}[\mathcal{A},\Phi]\rho_{\mathcal{R}}(g)^{-1}\,.
\end{equation}
Clearly, this modified equation has far more solutions than the bulk one. This can be illustrated by remembering that a small deformation in the trajectory of the line $x^\mu \to x^\mu+\delta x^\mu$ corresponds to an integrated insertion of $\rho_{\mathcal{R}}(F_{\mu\nu})\dot{x}^\mu\delta x^\nu$. But curiously we find that for a generic Lie algebra $\mathfrak{g}$ and representation $\mathcal{R}$, there are actually many inequivalent ways of inserting the field strength $\mathcal{O}^{(i)}(F_{\mu\nu})\in \text{End}(\mathcal{R})$, where $i=0,...,\Monedim-1$ with $\Monedim$ the number of non-zero Dynkin labels for $\mathcal{R}$. 
All these additional exotic operators have similar properties to the standard deformation $\mathcal{O}^{(0)}(F_{\mu\nu}) = \rho_\mathcal{R}(F_{\mu\nu})$, i.e. at least in weakly-coupled gauge theory their scaling dimension is close to 2, but for algebras of sufficiently large dimension and representations of sufficiently high charge there can be arbitrarily many of them. For example, if we choose an $SU(N)$ gauge theory, the operators $\mathcal{O}^{(i)}(X)$ for $X\in \mathfrak{su}(N)$ can be defined by
\begin{equation}
\label{eq:O_i_def}
    \mathcal{O}^{(i)}(X) = \Big(\rho_{\mathcal{R}}(T^a)\frac{\partial}{\partial X^a}\Big)^{i+1}\tr(X^{i+2})\,, 
\end{equation}
where $i\in\{0,...,\mathfrak{m}_\mathcal{R}-1\}$ and $T^a$ are the standard generators of $\mathfrak{su}(N)$. {The indices are raised and lowered with the Killing form $k_{ab}=2\tr_N(T_aT_b)$.}

Although these operators appear in any theory of gauge fields (including the Standard Model of particle physics), making inferences about them in full generality is beyond the scope of current techniques. Nevertheless, in simplified models with enhanced symmetry exact results may be derived.

\section{The spectrum of the $\frac{1}{2}$-BPS Wilson line in representation $\mathcal{R}$\label{sec:spectrum} }
We consider $\mathcal{N}=4$ SYM theory and focus on the most symmetric line operator \cite{Maldacena:1998im} extended along the $t$ axis
\begin{equation}
    \mathcal{W} = \mathds{P}\exp\int_{-\infty}^\infty  \big(iA_t+\Phi^6\big)dt\,,
\end{equation}
which preserves the algebra $\mathfrak{osp}(4^*|4)$ with half of the bulk supersymmetries in addition to the bosonic part $\mathfrak{sl}(2)\oplus\mathfrak{su}(2)\oplus \mathfrak{sp}(4)$. Among the preserved bosonic symmetries is the one-dimensional conformal algebra, which allows for the definition of defect CFT  (DCFT) correlation functions to be \cite{Drukker:2006xg}
\begin{equation}
    \average{O_1(t_1)...O_n(t_n)} \coloneqq \frac{\average{
\text{tr}_\mathcal{R}\mathds{P}\big[O_1(t_1)...O_n(t_n)\mathcal{W}\big]}_{\mathcal{N}=4}}
{
\average{\text{tr}_\mathcal{R}\mathcal{W}}_{\mathcal{N}=4}}\,.
\end{equation}
This DCFT enjoys a useful parity symmetry \footnote{
We use a definite-parity basis, with parity-even operators real and parity-odd ones imaginary. Their two-point functions are then positive/negative definite and the OPE coefficients are real with $C_{O_1 O_2 O_3}=s_1s_2s_3 C_{O_2 O_1 O_3}$,
see \cite{Homrich:2019cbt,Qiao:2017xif}.
} which, from the point of view of the $\mathcal{N}=4$ SYM theory, is a combination of $\mathsf{C}$ (charge conjugation) and $\mathsf{P}$ (parity in the $t$ direction),  see \cite{Cavaglia:2023mmu,Gomis:2025gzb}.
If $\mathfrak{g}$ possesses a non-trivial  group of outer automorphisms $\mathfrak{s}$, and $\mathcal{R}$ is invariant under its action, the DCFT has an extra  $\mathfrak{s}$ symmetry.
%We can use bulk charge conjugation $\mathsf{C}$ (which can be the identity operation for some $\mathfrak{g}$) in combination with parity $\mathsf{P}$ along the $t$ axis and the $R$-symmetry $\mathsf{R}$ that sends $\Phi^6 \to -\Phi^6$ to define an operation $\mathcal{P}=\mathsf{C}\mathsf{P}\mathsf{R}$, which acts as a defect parity symmetry for any Wilson line operator. Additionally, $\mathsf{C}$ acts as a $\mathbb{Z}_2$ symmetry for defect operators in self-dual representations. 

\subsection{Supersymmetric operators and associated structures\label{subsec:index}}
In this section we describe operators that preserve additional supersymmetry and the algebraic structures they give rise to.
%which makes their behaviour more constrained. 
First we direct our focus to the  $\textit{half-BPS}$ $D_k$ multiplet \footnote{See \cite{Ferrero:2023znz} for a review of the relevant supermultiplets.}. Their superprimaries are spinless with $\Delta=k$, transform in the $[0,k]$ representation of $\mathfrak{sp}(4)$ and are absolutely protected from recombination with other multiplets.
%A very concrete construction of them is given 
They can be constructed by considering equation (\ref{eqn_gauge_line_general}) for a single letter $X\in\mathfrak{g}$:
\begin{equation}
    \mathcal{O}_{line}(gX g^{-1}) = \rho_R(g)\mathcal{O}_{line}(X)\rho_R(g)^{-1}\,,
\end{equation}
whose solutions form a graded ring $\mathcal{M}= \oplus_{k\geq 0}\mathcal{M}_k$, where the weight-$k$ component consists of operators with fixed homogeneity $\mathcal{O}_k(\alpha X) = \alpha^k\mathcal{O}_k(X)$. Then the superprimaries of  $D_k$ are simply $\mathcal{O}_k(u^I \Phi^I)$ with $u^I$ an $\mathfrak{so}(5)$ null-vector and $\mathcal{O}_k\in\mathcal{M}_k$. In particular, for the weight one case we may introduce an explicit basis $\mathcal{O}_1^{(i)}(X) = \mathcal{O}_{1a}^{(i)}X^a$, where we take $\mathcal{O}_1^{(0)}(X) = \rho_\mathcal{R}(X)$.
Since $D_k$ multiplets cannot recombine, their number is coupling-independent and can be counted in the free theory, and a Hilbert series can be readily computed
\begin{equation}
    H_{\mathcal{R}}(x) = \sum_{k\geq 0}\dim(\mathcal{M}_k)x^k\,.
\end{equation}
The algebraic structure of the ring $\mathcal{M}$ is highly non-trivial for generic $\mathfrak{g}$ and $\mathcal{R}$. It is generally non-commutative \footnote{
It is commutative exactly when the weight-space decomposition of $\mathcal{R}$ is multiplicity-free. For $\mathfrak{su}(N)$, this leaves only the k-th symmetric and a-th antisymmetric representations.
}
and the generators satisfy an intricate web of relations (the ring has been studied in the mathematics literature \cite{Kirillov_family,Hausel_avatars}). In the bulk the analogous operators would be part of the Coulomb ring $\mathbb{C}[\mathfrak{g}]^G$ (for $\mathfrak{su}(N)$ freely generated by $\tr(X^2),...,\tr(X^N)$), but on the line $\mathcal{M}$ is a free $\mathbb{C}[\mathfrak{g}]^G$-module of rank $\sum_\mu (m_\mu)^2$, where $m_\mu$ denotes the multiplicity of the weight $\mu$ in $\mathcal{R}$.
 %The non-commutativity of the algebra is unique to one-dimensional systems;
 Non-commutativity is special to one dimension; consider an arbitrary correlator containing a pair of half-BPS operators with identical polarizations and take $\lim_{t_1\to t^{\pm}}\mathcal{O}_k(t_1,u)\mathcal{O}_l(t,u)$. This limit is non-singular and defines a new insertion $\mathcal{O}^{\pm}_{k+l}(t,u)\in \mathcal{M}_{k+l}$, but the answer depends on whether we approached $t$ from above or below. 
 The half-BPS chiral ring described here can be deformed, through a cohomological construction, into the topological sector of 
\cite{Drukker:2009sf,Liendo:2016ymz,Giombi:2018qox}, introducing a nontrivial \(g_{YM}^2\)-dependence.

Another class of operators are the $\textit{quarter-BPS}$ multiplets $B_1[a,b]$. Unlike their half-BPS cousins, these multiplets undergo recombinations and consequently their number is harder to determine. In the free theory there are two letters among the scalars, $X,Y\in \mathfrak{g}$, from which they can be constructed, therefore we must solve
\begin{equation}\label{quarterBPS_eqn}
    \mathcal{O}_{1/4}^{free}(gXg^{-1},gYg^{-1}) = \rho_\mathcal{R}(g)\mathcal{O}_{1/4}^{free}(X,Y)\rho_\mathcal{R}(g)^{-1}\,,
\end{equation}
but at one-loop many of these operators get an anomalous dimension.
The space of solutions to equation \eqref{quarterBPS_eqn} carries an action of
$\mathbb C^* \times SU(2)$. The $\mathbb C^*$ factor dilates
$X$ and $Y$ simultaneously, while $SU(2)$ rotates them as a doublet.
Accordingly, this space decomposes 
$
\mathcal M_{\frac14}^{free}
=
\bigoplus_{k,n\geq 0}
\mathcal M^{free}_{\frac14, k,n},
$
where $k$ denotes the $\mathbb{C}^*$ weight and $n$ labels the $n+1$ dimensional representation of $SU(2)$.
%representation.i.e.~$\mathcal{M}^{\frac{1}{4}} = \oplus_{m\geq n}\mathcal{M}^{\frac{1}{4}}_{m,n}$ with $a=2n, b=m-n$ \CM{sl2 and map}. 
Notice that the quantum numbers of a chiral ring operator are related to those of its $B_1[a,b]$ ancestor as $k=a+b$ and $n=b$.
To figure out which of these operators remain protected in the interacting theory one should compute the cohomology of the relevant supercharge $\mathfrak{Q}$ \cite{Cachazo:2002ry,Grant:2008sk,Chang:2013fba}. For example, the number of $B_1[2,0]$ multiplets in the free theory is $\Monedim^2$ where $\Monedim=\dim(\mathcal{M}_1)$ and they may be written as
\begin{equation}
    \mathcal{O}^{(i,j)}_{(2,0)}(X,Y):= \mathcal{O}^{(i)}_1(X)\mathcal{O}^{(j)}_1(Y)-\mathcal{O}^{(i)}_1(Y)\mathcal{O}^{(j)}_1(X)\,,
    \label{eq:B1[2,0]}
\end{equation}
with $i,j=0,\dots,\Monedim-1$.
Among these the combination $ \mathcal{O}^{(0,i)}_{(2,0)}(X,Y)+\mathcal{O}^{(i,0)}_{(2,0)}(X,Y)={2}\mathcal{O}^{(i)}_1([X,Y])$
and, for $g_{YM}>0$, it is recognized to be $\mathfrak{Q}$-exact. 
The remaining $\Monedim(\Monedim-1)$ are protected since there is no operator they can recombine with.
Half of these are even and half are odd under one-dimensional parity, the associated supermultiplets will be denoted as $B_1[2,0]^{\pm}$.

\section{Exact result for the leading beta-function}\label{sec:breaking_conf_symm}
In this section, we investigate what happens to the DCFT when deformed by  
\begin{equation}
\label{deformation}
\zeta^{\mathcal{I}}
\int dt \,D_{\mathcal{I}}(t)\,,
\qquad
D_{(i,I)}= \tfrac{2\pi}{g_{YM}}
O_{1a}^{(i)}\,\Phi^a_I+...\,,
\end{equation}
where $\mathcal{I}=(i,I)$, $i=0,\dots,\Monedim-1$ and $I=1,\dots,5$.  
 The prefactor in \eqref{deformation} fixes the leading-order normalisation
$\average{D_{\mathcal{I}}(0)D_{\mathcal{J}}(\infty)}
=g_{\mathcal{I}\mathcal{J}}=\mathsf{g}_{ij}\delta_{IJ}$,
which we also impose at higher orders in $g_{YM}$.
%The prefactor in \eqref{deformation} fixes the normalisation $\average{D_{\mathcal{I}}(0)D_{\mathcal{J}}(\infty)}=g_{\mathcal{I}\mathcal{J}}=\mathsf{g}_{ij}\delta_{IJ}$ at leading order in $g_{YM}$. At higher loops we normalise these operators to keep this form of the two-point function.
These operators have protected conformal dimension $\Delta=1$ for any value of the Yang-Mills coupling but can acquire an anomalous dimension once the deformation is turned on \footnote{Because of unitarity bounds the undeformed theory does not have relevant operators.}.
Since the perturbation \eqref{deformation} -- with the exception of the direction associated to the super-displacement operator -- breaks all the supersymmetries, we cannot use the mechanism of multiplet recombinations, see e.g.~\cite{Green:2010da}, to understand which direction acquires an anomalous dimension. 
We therefore compute the beta-function for the couplings $\zeta^{\mathcal{I}}$ directly by relating it to integrated higher point functions of $D_{\mathcal{I}}$ operators 
using the techniques of \cite{Komargodski:2016auf,Behan:2017emf,Behan:2017mwi}. 
As these operators have $\Delta=1$ and their three-point functions vanish by $\mathfrak{sp}(4)$ R-symmetry in the undeformed theory, the first non-trivial correction to the beta-function is of order $\zeta^3$, explicitly 
\begin{equation}
\label{betafunction1}
\beta^{\mathcal{I}}:=\frac{d \zeta^{\mathcal{I}}}{d \log(1/a)}=
-\frac{1}{3}\zeta^{\mathcal{J}} \zeta^{\mathcal{K}} \zeta^{\mathcal{L}} \,\,
\delta_{\mathcal{J}}C_{\mathcal{K}\mathcal{L}}^{\,\,\,\,\,\,\,\,\mathcal{I}}+\dots\,,
\end{equation}
where $a$ is a short-distance cutoff and
\begin{equation}
\label{C_IJKL_int_formula}
\frac{\delta_{\mathcal{I}}C_{\mathcal{J}\mathcal{K}\mathcal{L}}}{t_{32}t_{31}t_{21}}=
\int dt\,\langle D_{\mathcal{I}}(t)D_{\mathcal{J}}(t_1)D_{\mathcal{K}}(t_2)D_{\mathcal{L}}(t_3) \rangle\,,
\end{equation}
with $t_1<t_2<t_3$, is the variation of the three-point function. 
To produce a finite quantity from this integral, one needs to regularise it by restricting $|t-t_i|>a$ and subtracting the $1/a$ divergences originating from the exchange of the identity operator in the OPEs of two $D$
operators.
% Mention math ordering.
Finally, indices are raised with the inverse of the metric $g_{\mathcal{I}\mathcal{J}}$.

In the following we show that, using the constraints of superconformal symmetry on the four point function of $D_1$-type operators \cite{Liendo:2018ukf}, the integral \eqref{C_IJKL_int_formula} can be expressed in terms of the OPE coefficients associated to the exchange of $B_1[2,0]^+$ multiplets. 
Explicitly, \eqref{C_IJKL_int_formula} 
%the variation of the OPE coefficients 
reduces to \footnote{Notice that
$(C^{\,2}_{B_1[2,0]})_{(ij)k\ell}=2(C^{\,2}_{B_1[2,0]^+})_{ijk\ell}$ where $A_{(ij)}=A_{ij}+A_{ji}$.}
\begin{equation}
\begin{aligned}
\label{deltaCtext}
\delta_{\mathcal{I}}C_{\mathcal{J}\mathcal{K}\mathcal{L}}=&
{-}2\,\delta_{IJ}\delta_{KL}
\bigl(C^{\,2}_{B_1[2,0]^+}\bigr)_{ij,k\ell}\\
&
{-}
2\,\delta_{IK}\delta_{LJ}
\bigl(C^{\,2}_{B_1[2,0]^+}\bigr)_{ik, \ell j}\\
&
{-}
2\,\delta_{IL}\delta_{JK}
\bigl(C^{\,2}_{B_1[2,0]^+}\bigr)_{i\ell, jk}\,,
\end{aligned}
\end{equation}
where
\begin{equation}
\label{C2Xdefinition}
\bigl(C^{\,2}_X\bigr)_{ij,k\ell}=
\sum_{\mathcal{O},\mathcal{O}' \\
\text{of type $X$}}
C_{ij}^{\,\,\,\,\,\mathcal{O}}C_{k\ell}^{\,\,\,\,\,\mathcal{O}'}g_{\mathcal{O}\mathcal{O}'}\,,
\end{equation}
and $C_{ij}^{\,\,\,\,\,\mathcal{O}}$ are OPE coefficients.
{By specializing the index $\mathcal{I}$ in equation \eqref{deltaCtext} to the direction of the displacement super-multiplet $(0,I)$, the condition $\delta_{(0,I)}C_{\mathcal{J}\mathcal{K}\mathcal{L}}=0$ implies  a universal constraint, in the spirit of \cite{Gabai:2025zcs,Girault:2025kzt,Drukker:2025dfm,Belton:2025ief}, on the quarter-BPS chiral ring} \footnote{
Equivalently this can be rewritten as $C_{0i\mathcal{O}}+C_{i0\mathcal{O}}=0$
$\forall$ $\mathcal{O}$ of type $B_1[2,0]$.
}:
\begin{equation}
\label{eq:constrain_for_displacement}
C_{0i}^{\,\,\,\,\,\mathcal{O}}=0\,,
\qquad
\forall\,\,\, \mathcal{O}\,\,\text{of type $B_1[2,0]^{+}$}. 
\end{equation}
As explained by $\mathfrak{Q}$-cohomology below \eqref{eq:B1[2,0]}, this holds for the half-BPS Wilson line in $\mathcal{N}=4$ SYM once we turn on the Yang-Mills coupling. 
The formula for the beta-function is obtained by inserting \eqref{deltaCtext} in \eqref{betafunction1} %and using the symmetries of $C^{\,2}_{B_1[2,0]^{+}}$
\begin{equation}
\label{betafunctionwithC}
\beta^{(i,I)}=
{+}2 \zeta^{(\ell,I)}\,(\zeta^{j}\cdot \zeta^{k})\,
\bigl(C^{\,2}_{B_1[2,0]^{+}}\bigr)_{jk,\ell i'}\,\mathsf{g}^{i'i}+O(\zeta^5)\,,
\end{equation}
where $\cdot$ denotes the scalar product of five-vectors.
At this order in $\zeta$, this is a gradient flow. We show by
elementary methods that the half-BPS line is locally IR attractive
for arbitrary $g_{YM}$. The proof, given in the Supplemental
Materials, uses the positivity properties of $C^{\,2}_{B_1[2,0]^+}$
to show that the distance to the BPS locus decreases towards the
infrared for sufficiently small exotic deformations, establishing
their marginal irrelevance.
%At this order in  $\zeta$ this is a gradient flow. We can show by elementary methods that the half-BPS line is IR attractive in a neighbourhood surrounding it in the $\zeta$-plane for arbitrary $g_{YM}$. The proof, discussed in the  Materials, shows that the distance of an arbitrary point to the BPS-locus shrinks under RG flow and relies on the positivity properties of $C^{\,2}_{B_1[2,0]^{+}}$: sufficiently small exotic deformations flow back to the half-BPS locus in the infrared and are marginally irrelevant.
%The analysis does not assume the operator is a Wilson line and applies to all $\mathfrak{osp}(4^*|4)$ invariant line-defects in $\mathcal{N}=4$ SYM theory, demonstrating they must all be infrared stable \footnote{The argument does not exclude the possibility of a non-trivial defect conformal manifold. Its existence would require relations in the quarter-BPS chiral ring.}.
The analysis applies to all $\mathfrak{osp}(4^*|4)$-invariant line defects in $\mathcal{N}=4$ SYM, not just Wilson lines, demonstrating their infrared stability \footnote{ A nontrivial defect conformal manifold is not excluded, but would require relations in the quarter-BPS chiral ring.}.

In the next section we compute, with two different methods, the tensor $C^{\,2}_{B_1[2,0]^{+}}$ at leading order in the Yang-Mills coupling. 
\begin{figure}
    \centering
    $\mathsf{g}_{ij}=$\TraceGij $\quad \mathsf{b}_{ijk\ell}=$\TraceBijkl 
    $\quad \mathsf{x}_{ijk\ell}=$\TraceXijkl
    \caption{
    Trace diagrams for the structures $\mathsf{g}_{ij}$,
$\mathsf{b}_{ijk\ell}$ and $\mathsf{x}_{ijk\ell}$.
Indices label insertions, outer circles denote traces over
$\mathcal{R}$, and red lines contract adjoint indices.
The blue dot in $\mathsf{x}_{ijk\ell}$ represents the
%four-index
vertex $f^{ab}{}_e f^{cde}$.}
    %Trace diagrams for the tensor structures $\mathsf{g}_{ij},\mathsf{b}_{ijk\ell}$ and $\mathsf{x}_{ijk\ell}$.
 %   The indices label the insertions,
    %The three independent tensor structures $\mathsf{g}_{ij},\mathsf{b}_{ijk\ell}$ and $\mathsf{x}_{ijk\ell}$ can be associated to the diagrams generating the corresponding contraction of colour indices. 
 %   {the outer circle is a trace over the representation, the red lines are contractions of adjoint indices,
   % the blue dot in $\mathsf{x}_{ijk\ell}$ represents the four-index vertex $f^{ab}{}_ef^{cde}$.}
    \label{fig:tensor_struct}
\end{figure}
\subsection{Integrated four-point function at finite coupling\label{subsec:integrated_corr}}

Super-conformal symmetry implies that the correlation function of four $D_1$-type operators can be written as \cite{Liendo:2018ukf}
%\cite{Artico:2024wut}  
\begin{equation}
\begin{aligned}
\langle &D_1^{(i)}(1)D_1^{(j)}(2)D_1^{(k)}(3)D_1^{(\ell)}(4) \rangle = \\
&=\,(13)(24)
\left(\mathbb{F}_{ijk\ell}-\partial_\chi\left(\tfrac{(\chi-y_1)(\chi-y_2)}{\chi^2}
f_{ijk\ell}(\chi)\right)
\right)
\end{aligned}
\label{eq:4pt_both_formulas}
\end{equation}
where the conventions are the same as in \cite{Ferrero:2021bsb}.
In this equation  $t_1<t_2<t_3<t_4$ and $\chi=\tfrac{t_{12}t_{34}}{t_{13}t_{24}}$, so that in the integral  \eqref{C_IJKL_int_formula} 
 the labels $i,j,k,\ell$ must  be permuted to match the ordering
of the insertion points.
%we have to rearrange the $i,j,k,\ell$ labels according to the corresponding points.
The result in \eqref{eq:4pt_both_formulas} is useful to recognize that, up to the $\mathbb{F}_{ijk\ell}$ contribution, we are integrating a total derivative \footnote{
The integral thus reduces to boundary terms determined by OPE
coefficients. See  \cite{Chen:2026vml} for a similar mechanism.}.
%This implies that the integration produces boundary terms that we can evaluate in terms of OPE coefficients; a similar mechanism takes place in \cite{Chen:2026vml}.
%
Additionally, crossing symmetry reads
\begin{equation}
\label{eq:crossing}
(1-\chi)^2f_{ijk\ell}(\chi)+
\chi^2 f_{jk\ell i}(1-\chi)=0\,.
\end{equation}
%In each region of integration the integrand is a total derivative, and the integration gives boundary terms which correspond to evaluating the function $f(\chi)$  for $\chi$ close to zero or one.
%These terms are completely captured by the OPEs, and in the $a\rightarrow 0$ limit only the exchange of the identity, $D_2$-type multiplets and $B_1[2,0]$  multiplets contribute
In each integration region, the integrand is a total derivative,
giving boundary terms obtained by evaluating $f(\chi)$ near
$\chi=0,1$. These are completely determined by the OPEs, and in the $a\rightarrow 0$ limit only the exchange of the identity, $D_2$ and $B_1[2,0]$  multiplets contribute
\footnote{
The limits $g_{YM}^2\to0$ and $a\to0$ do not commute
because of terms $a^{\Delta(g_{YM}^2)-1}$ with $\Delta(0)=1$.
This is resolved by including perturbative mixing between
$\Phi^I$ and $\Phi^6$; see also~\cite{Cavaglia:2022qpg,Gabai:2025zcs}.
}.
%\footnote{
%
%Some care is needed as the  limit   $g^2_{YM}\rightarrow 0$ does not commute with removing the short-distance regulator because of terms of the form  $a^{\Delta(g^2_{YM})-1}$ with $\Delta(0)=1$. 
%This issue is resolved by considering a perturbative mixing between $\Phi^I$  and $\Phi^6$.
 %   This mechanism is essentially discussed in \cite{Cavaglia:2022qpg,Gabai:2025zcs}.
%
%}. 
The contribution of the identity is completely cancelled by counterterms and the one of $D_2$ multiplets by the contribution from $\mathbb{F}$. 
This cancellation relies on the total symmetry of $F_{ijk\ell}$,
which follows from chiral-ring arguments~\cite{Artico:2026tba}.
%For this cancellation to take place we use that $\mathbb{F}_{ijk\ell}$ is totally symmetric in its indices, as can be shown from half-BPS chiral ring arguments \cite{Artico:2026tba}.
The final result, after stripping off the R-symmetry polarization vectors, is equation \eqref{deltaCtext}.

%------
\section{The four-point function of $D_1$ operators at weak coupling\label{sec:4pt_NLO}}

This section is dedicated to the calculation of the four-point function \eqref{eq:4pt_both_formulas} up to next-to-leading order in $g_{YM}^2$.
%This section is dedicated to the perturbative calculation of the four-point function \eqref{eq:4pt_both_formulas}
% up to next-to-leading order in the perturbative expansion for small $g_{YM}^2$.
%\eqref{C_IJKL_int_formula} 
The result is universal and holds for any gauge algebra $\mathfrak{g}$ and representation $\mathcal{R}$. The dependence on these data is via the tensors $\mathsf{g}_{ij},\mathsf{b}_{ijk\ell}$ and $\mathsf{x}_{ijk\ell}$. Each of these tensors can be associated to a trace diagram generating the corresponding colour index contraction, see Fig.~\ref{fig:tensor_struct}. 
Details are found in the Supplemental Materials.
From this result we can extract the 
leading order contribution in an expansion in $g^{2}_{YM}$ to the beta-function \eqref{betafunctionwithC}.
Because of multiplet recombination, to extract $C^2_{B_1[2,0]^+}$ at a given order we need the correlator \eqref{eq:4pt_both_formulas} at the next order.\\
\begin{table}%[H]
    \centering
    \caption{
  Feynman diagrams contributing to $f^{(1)}_{ijk\ell}(\chi)$. The first three involve $4d$ bulk integrals. In the last row, orange dots along the Wilson line mark possible gluon
attachment points, giving $1d$ integrals.
%Feynman diagrams contributing to $f^{(1)}_{ijk\ell}(\chi)$.
%    The first three are bulk diagrams and correspond to $4d$ integrals.
 %   In the last row, the orange dots along the Wilson line indicate where the gluon can connect and generate a $1d$ integral.
    }
    \begin{tabular}{lc}
        \hline
        Self-energy & \scaleddiagram{\DefectSSSSSelfEnergyOne}\ \scaleddiagram{\DefectSSSSSelfEnergyTwo}\ \\[2ex]
        \hline
        Four-scalar vertex & \scaleddiagram{\DefectSSSSX}\ \\[2ex]
        \hline
        Gluon exchange & \scaleddiagram{\DefectSSSSH}\ \\[2ex]
        \hline
        Gluon to line & \scaleddiagram{\DefectSSSSYOne}\ \scaleddiagram{\DefectSSSSYTwo}\ \\[2ex]
        \hline
    \end{tabular}
    \label{table:Diagrams1111NLO}
\end{table}
%
%This section is dedicated to the perturbative calculation of $\mathcal{A}_{ijkl}(\chi,r,s)$ \RT{This has not been defined before. We should call it something else.}\DA{Proposal to write instead just $f_{ijkl}(x)$} up to next-to-leading order in the perturbative expansion for small $g_{YM}^2$ at finite $N$. The result is expressed as a function of constants that depend on the label of the operators and on the representation $\mathcal{R}$. We provide in Appendix \DA{add} the formulas to compute these coefficients. A clear link to the integrated correlator of equation \eqref{eq:Integral_after_OPE} is established by extracting the value of $\mathcal{I}_{ijkl}$ at leading order, as well as the anomalous dimensions of new operators at the unitarity bound.
%------------------------

%\vspace{1cm}
It is convenient to use
$\hat{\lambda}_{\mathfrak{g}}:=
\frac{g^2_{YM} h_{\mathfrak{g}}^{\vee}}{4 \pi^2}$ as expansion parameter and consider operators normalized as discussed below \eqref{deformation}. In the free theory, we find that the building blocks in  \eqref{eq:4pt_both_formulas} are
\begin{equation}
\label{F0andf0}
\begin{aligned}
\mathbb{F}^{(0)}_{ijk\ell}&=\mathsf{g}_{ij}\mathsf{g}_{k\ell}+\mathsf{g}_{i\ell}\mathsf{g}_{jk}+\mathsf{b}_{ijk\ell}\,,\\
f^{(0)}_{ijk\ell}&=\tfrac{\chi}{1-\chi}
\left((1-\chi) \mathsf{g}_{ij}\mathsf{g}_{k\ell}-\chi\,\mathsf{g}_{i\ell}\mathsf{g}_{jk}\right)\,.
\end{aligned}
\end{equation}
%where $\mathsf{g}_{ij}$ and $\mathsf{b}_{ijk\ell}$ are given in Eq.~\eqref{gijdef} and \eqref{eq:x_and_b}. \MS{to be fixed}
To compute the NLO correction, we follow the perturbative bootstrap approach of \cite{Artico:2024wut}. We first evaluate the $(13)(24)$
polarization from the diagrams in Table~\ref{table:Diagrams1111NLO},
then extract $f^{(1)}_{ijk\ell}(\chi)$ by solving a differential
equation and reconstruct the full four-point function using
Eq.~\eqref{eq:4pt_both_formulas}.
%To compute the NLO correction, we first focus on a Feynman-diagrammatic calculation of the polarization \((13)(24)\), from which we extract \(f_{ijkl}(\chi)\) by solving a differential equation. We then reconstruct the full four-point function using Eq.~\eqref{eq:4pt_both_formulas}. This procedure follows the perturbative bootstrap approach introduced in \cite{Artico:2024wut}. The Feynman diagrams contributing to the polarization under consideration are collected in table \ref{table:Diagrams1111NLO}.
The final result is remarkably simple:
\begin{equation}
\begin{aligned}
\label{F1andf1}
\mathbb{F}^{(1)}_{ijk\ell}&=  \tfrac{\pi^2}{6}\,\mathsf{x}_{ijk\ell}\,,\\
f^{(1)}_{ijk\ell}&=
\tfrac{\chi}{1-\chi}
 (u(\chi)-u(1-\chi))\,\mathsf{x}_{ijk\ell}\,,
\end{aligned}
\end{equation}
where the totally symmetric tensor $\mathsf{x}_{ijk\ell}$ comes from the colour contraction shown in Fig.~\ref{fig:tensor_struct} and $u(\chi)=\frac{\text{Li}_2(\chi)}{2}-\frac{\pi^2}{12} \chi$.
By expanding the correlator in super-conformal blocks, which has to be done carefully because of the exchange of operators near the unitarity bound, we obtain:
\begin{equation}
\label{C2andL}
\begin{aligned}
(C^2_{B_1[2,0]})^{(0)}_{ijk\ell}+\mathcal{L}_{ijk\ell}&=\tfrac{1}{2}(\mathsf{g}_{i\ell}\mathsf{g}_{jk}-\mathsf{b}_{ijk\ell})\,,
\\
\hat{\mathcal{L}}_{ijk\ell} &=\tfrac{1}{2} \mathsf{x}_{ijk\ell}\,,
\end{aligned}
\end{equation}
where we defined
\begin{equation}
\begin{aligned}
\mathcal{L}_{ijk\ell}&=
\lim_{g^2_{YM}\rightarrow 0} \sum_{\mathcal{O}}\frac{C_{ij \mathcal{O}}C_{k\ell\mathcal{O}}}{\Delta_{\mathcal{O}}-1}\,,
\\
\hat{\mathcal{L}}_{ijk\ell}&=
\lim_{g^2_{YM}\rightarrow 0} \sum_{\mathcal{O}}\frac{C_{ij \mathcal{O}}C_{k\ell\mathcal{O}}}{\hat{\lambda}_{\mathfrak{g}}}\,,
\end{aligned}
\end{equation}
with the sum running over super-primary operators near the unitarity bound $\Delta=1$.
Using the mechanism of multiplet recombinations, it can be shown that all operators $\mathcal{O}$ in the sum above satisfy $\Delta=1+\hat{\lambda}_{\mathfrak{g}}+\dots$ so that $\mathcal{L}=\hat{\mathcal{L}}$. 
In this way we can use \eqref{C2andL} to extract $(C^2_{B_1[2,0]})$.
We have validated this result by directly computing $(C^2_{B_1[2,0]})^{(0)}$ via summing over all the $B_1[2,0]$ multiplets which do not get anomalous dimensions. 
In the Supplemental Materials, we further check this result by recovering its weak-coupling limit from the one-loop beta-function for  deformations of the ordinary Wilson line. 

\section{Discussion \label{sec:conclusion}}
In this letter we have shown that Wilson lines admit a spectrum of
excitations far richer and more intricate than that of the bulk.
Further details and results will be reported in \cite{Artico:2026tba}.
For half-BPS Wilson lines in $\mathcal{N}=4$ SYM, many of these
operators share the quantum numbers of the super displacement\footnote{A similar multiplicity of displacement-type multiplets
occurs for half-BPS defects in the $6d$ $(2,0)$ theory; see \cite{Meneghelli:2022gps}.} and can be used to deform the defect.
 Their leading beta-functions are given non-perturbatively in $g_{YM}$ by OPE coefficients involving quarter-BPS chiral ring operators. This allows us to show that the half-BPS  Wilson line is rigid, i.e. IR stable, under the deformations associated with the exotic operators constructed here.
Interestingly, the argument applies to any half-BPS line in $\mathcal{N}=4$ SYM, with the possibility of having a defect conformal manifold if exotic operators satisfy extra relations in the quarter-BPS  chiral ring.
A natural extension is to compute the $\zeta^5$ term in the beta-function from integrated six-point correlators
and test whether the reduction to chiral-ring data
persists beyond cubic order.

%It would be useful to extend the non-perturbative analysis of this letter and compute the $\zeta^5$beta function by integrating  six-point correlators
%determine if it is still controlled by chiral ring data.

%In particular, we have shown that the half-BPS line in $\mathcal{N}=4$ SYM is stable under all the deformations generated by the exotic operators [37], with a leading beta-function fixed non-perturbatively by OPE coefficients of quarter-BPS chiral-ring operators.
% In particular,  \col{we have argued that the half-BPS line in $\mathcal{N}=4$ SYM is rigid under the deformations associated with the exotic operators constructed here} 

\begin{figure}
    \centering
    \includegraphics[width=0.8\linewidth]{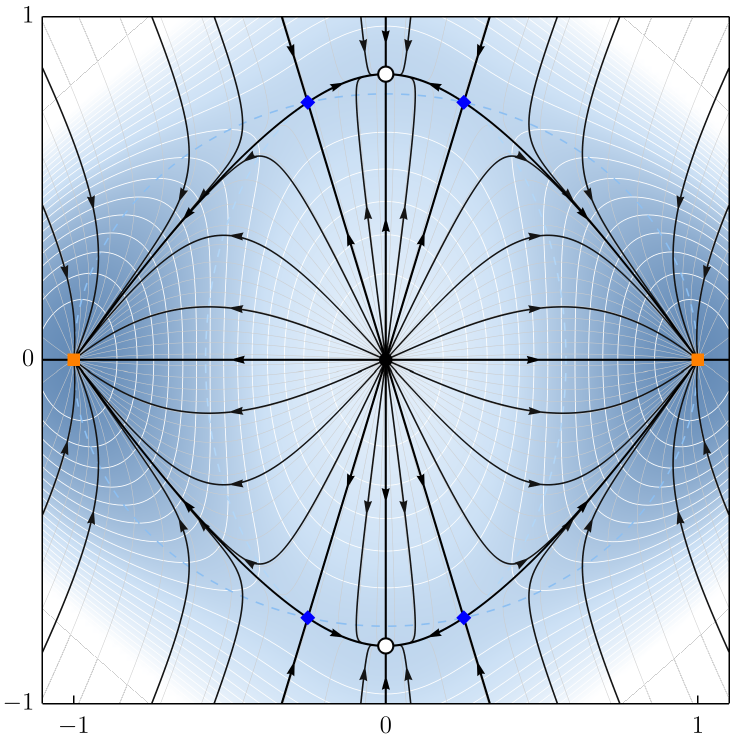}
    \caption{One-loop RG flow for the $SU(3)$ adjoint line, restricted to a
fixed scalar direction. The origin is the ordinary line.
Orange, blue and white points denote half-BPS, mixed and purely
exotic fixed points, respectively. See the Supplemental Materials
for details.
%RG flow for the adjoint representation of  $SU(3)$ projected on a plane. 
 %   The origin is the ordinary Wilson line, the orange dots correspond to $1/2$-BPS lines, the blue dot to mixed exotic-ordinary fixed points, and the white dots to pure exotic lines. See Supplemental Materials for more details.
 }
    \label{fig:rg_flow}
\end{figure}

Similar deformations generate multidimensional RG flows from ordinary
Wilson lines in $\mathcal{N}=4$ SYM, generalizing those studied in
\cite{Polchinski:2011im,Beccaria:2022bcr}. There are $\Monedim$
operators transforming as $\mathfrak{so}(6)$ vectors, which are relevant in the DCFT, at least for small Yang-Mills coupling \footnote{
These operators may become more relevant at stronger coupling.
For the fundamental ordinary line in the planar limit, their
scaling dimension tends to zero at large 't Hooft
coupling~\cite{Beccaria:2017rbe}.
}.
Our preliminary one-loop analysis reveals previously unknown non-supersymmetric fixed points in addition to the
half-BPS Wilson line.
The simplest example is the $SU(3)$ adjoint line,
where $\Monedim=2$ gives twelve deformation parameters.
Fig.~\ref{fig:rg_flow} shows the flow restricted to the plane where the two (ordinary and exotic) six-vectors are parallel.
A systematic investigation  of these RG flows and their newly identified fixed points 
is an important step towards understanding the space of conformal line defects. Higher-order
perturbation theory could test their persistence and stability, while the numerical bootstrap could constrain their spectra and OPE coefficients. A complementary approach is to explore large-charge limits with large Dynkin labels \cite{Aharony:2022ntz}, in which the path integral may be amenable to saddle-point methods.

For generic representations, the multiplicity of displacement-type
multiplets matches the number of Dynkin labels, suggesting a
connection between defect deformations and continuous variations
of these labels. While it may be possible to analytically
continue the DCFT data in the Dynkin labels and define universal
structures analogous to those in
\cite{Chang:2024zqi,Bonetti:2025kan}, our local exotic deformations
do not generate a conformal manifold \footnote{
For a recent study of the relation between continuous families
of CFTs and conformal manifolds see \cite{Komatsu:2025cai}.
}.
An enlarged defect theory with additional nonlocal degrees of
freedom, inspired by the long-range/short-range crossover
\cite{Behan:2017emf}, might nevertheless admit a continuous family
of conformal defects parametrized by these labels, without
contradicting the original Wilson line's local infrared stability.

Another question concerns the spectrum of Wilson lines for groups
of large rank. At strong 't Hooft coupling $\hat{\lambda}_{\mathfrak{g}}$, these lines correspond
to extended objects in a weakly coupled string theory in
$\mathrm{AdS}_5$.
%Another question concerns the spectrum of Wilson lines for groups of large rank. At large 't Hooft coupling $\hat{\lambda}_{\mathfrak{g}}$ this limit is described as a weakly coupled string theory in $AdS_5$ and the Wilson lines correspond to extended objects propagating in this gravitational dual. 
The novelty of these configurations is the additional presence of open string states, therefore we expect an intimate relation (if not outright equivalence) between the Hilbert spaces $\mathcal{H}_{exotic}$ and $\mathcal{H}_{open}$. For example, the fundamental representation has a dual $F1$ string and as $\hat{\lambda}_{\mathfrak{su}(N)}\to \infty$ all of the states are generated by taking symmetric products of a single $D_1$ multiplet, which is interpreted as describing the embedding of the string worldsheet. For representations with more boxes the dual is given by $D3/D5$ branes \cite{Gomis:2006sb}, so the exotic, protected multiplets at $\Delta=1$ should acquire a geometric interpretation as their moduli. A clarification of this point should give a prediction for the spectrum of a Wilson line in an arbitrary representation as $\hat{\lambda}_{\mathfrak{su}(N)}\to \infty$. These claims could be further tested by extending localisation methods to observables in the topological sector, or by analysing the supersymmetric index and the associated ring. Either approach would offer a window into the strong-coupling regime.\\

\begin{acknowledgments}
The authors thank Julien Barrat, Marisa Bonini, Luca Griguolo and Philine van Vliet for useful discussions. DA is grateful to Adam Chalabi and Silvia Penati for the conversations during his stay at the Isaac Newton Institute for Mathematical Sciences, Cambridge, during the workshop \textit{Quantum Field theory with boundaries, impurities, and defects}.
%---
DA and CM are  supported by the MUR PRIN contract 2022N9CTAE “Constraining strongly coupled quantum field theories using symmetry”.
RT is supported by University of Parma through  the action Bando di Ateneo 2023 per la ricerca.
All authors acknowledge support from INFN through the GAST research project.
\end{acknowledgments}


%\bibliography{letter/letter}  
\putbib

\end{bibunit}

\clearpage
\onecolumngrid

\appendix
\begin{center}
   \large{ \textbf{Supplemental Materials to New Exotic Operators \\ in the Spectrum of Wilson Lines in General Representations}}\\[.3cm]
   \normalsize {Daniele Artico, Carlo Meneghelli, Michele Savi, and Rudolfs Treilis}
\end{center}
\setcounter{page}{1}
\begin{bibunit}
\section{I: Hilbert series}
The Hilbert series $H_\mathcal{R}(x) = \sum_{k\geq 0}\dim(\mathcal{M}_k)x^k$ counts the number of weight-$k$ half-BPS operators. The coefficients $\dim(\mathcal{M}_k)$ are integers, which are also equal to the number of times the trivial representation $\mathds{1}$ appears in the tensor product $\mathcal{R}\otimes \bar{\mathcal{R}}\otimes\text{Sym}^k(\text{Adj})$. An efficient method to compute this series is to use the orthogonality of group characters $\chi_\mathcal{R}(g)$, from which it follows
\begin{equation}
    \dim(\mathcal{M}_k) = \int_G d\mu_g \chi_\mathcal{R}(g)\chi_{\bar{\mathcal{R}}}(g)\chi_{\text{Sym}^k(\text{Adj})}(g)\,,
\end{equation}
where $d\mu_g$ is the Haar measure. The generating series over characters of symmetric powers of the adjoint representation can be computed and after using the Weyl integration formula we arrive at
\begin{equation}
    H_\mathcal{R}(x) = (1-x)^{-r}\frac{1}{|W|}\oint_{\mathbb{T}^r}\prod_{a=1}^r\frac{d z_a}{2\pi iz_a} \chi_\mathcal{R}(z)\chi_{\bar{\mathcal{R}}}(z)\prod_{\alpha\in \Delta}\frac{1-e^\alpha(z)}{1-x e^{\alpha}(z)}\,,
\end{equation}
    where $r$ is the rank of $\mathfrak{g}$, $W$ is the Weyl group, $\chi_{\mathcal{R}}(z)$ is the character evaluated on an element of the maximal torus $\mathbb{T}^r$ and the product is over all the roots $\alpha\in \Delta$ of the Lie algebra. The Hilbert series is a rational function of $x$ and it has the form
    \begin{equation}
        H_{\mathcal{R}}(x) = \frac{P_{\mathcal{R}}(x)}{\prod_{a=1}^r(1-x^{d_a})}\,,
    \end{equation}
where the Coulomb ring $\mathbb{C}[\mathfrak{g}]^G \cong \mathbb{C}[C_{d_1},...,C_{d_r}]$ is freely generated by the Casimirs $C_{d_a}$ of weight $d_a$ and $P_{\mathcal{R}}(x)$ is a polynomial. Consequently the entirety of the information in the Hilbert series is contained in the finite number of coefficients of the polynomial. We have managed to figure out the cases of general representation for $\mathfrak{su}(2)$ and $\mathfrak{su}(3)$, which are given by
\begin{equation}
    P^{\mathfrak{su}(2)}_{[k]}(x) = \sum_{l=0}^k x^l\,,\qquad P_{[k_1,k_2]}^{\mathfrak{su}(3)}(x) = N_{k_1}(x)N_{k_2}(x)+x^3N_{k_1-1}(x)N_{k_2-1}(x)\,,\quad N_{k}(x) = \sum_{\substack{a,b,c\geq 0\\ a+b+c=k}} x^{b+2c}\,.
\end{equation}
We did not manage to find general formulas for other algebras, but expect that the coefficients of the polynomial likely have some combinatorial interpretation. A possible approach for generalization might be to construct generating series over representations, which are given by
\begin{align}
    G^{\mathfrak{su}(2)}(y;x) = \sum_{k=0}^\infty y^k P^{\mathfrak{su}(2)}_{[k]}(x) = \frac{1}{(1-y)(1-xy)}\,,\\
    G^{\mathfrak{su}(3)}(y_1,y_2;x) = \sum_{k_1,k_2=0}^{\infty}y_1^{k_1}y_2^{k_2}P^{\mathfrak{su}(3)}_{[k_1,k_2]}(x) = \frac{1+y_1y_2x^3}{\prod_{i=1}^2(1-y_i)(1-xy_i)(1-x^2y_i)}\,.
\end{align}
Although we do not know the exact Hilbert series for generic representations, in the case of $\mathfrak{su}(N)$ rank-$k$ symmetric and anti-symmetric they can be easily identified
\begin{equation}
    P^{\mathfrak{su}(N)}_{\text{Sym}^k(\Box)}(x) = \prod_{a=1}^k\frac{1-x^{N+a-1}}{1-x^a}\,,\qquad P^{\mathfrak{su}(N)}_{\wedge^k(\Box)}(x) = \prod_{a=1}^k\frac{1-x^{N-k+a}}{1-x^a}\,.
\end{equation}
Observe that $\lim_{N\to \infty} P^{\mathfrak{su}(N)}_{\text{Sym}^k(\Box)}(x) = \lim_{N\to \infty}P^{\mathfrak{su}(N)}_{\wedge^k(\Box)}(x)= \prod_{a=1}^k(1-x^a)^{-1}$, so for fixed $k$ the half-BPS spectra of symmetric and anti-symmetric lines become identical in the large-$N$ limit. Finally, we remark that the linear term $H_{\mathcal{R}}(x)=1+\Monedim x +...$ has a universal form encoding the number of non-zero Dynkin labels.

\section{II: Colour factors}

Here we present the colour structures that appear, discuss their properties, and explain how they are evaluated explicitly in the case of $\mathfrak{g}=\mathfrak{su}(N)$.
We denote by $ O_{1a}^{(i)}\,\in\,\text{End}(\mathcal{R})$, $i=0,\dots,\Monedim-1$ a basis of linearly independent, Hermitian solutions of the equation
\begin{equation}
\label{equivarianceAppendix}
[\rho_{\mathcal{R}}(T_a),\,O_{1b}^{(i)}]
={i}f_{a b}^{\,\,\,\,c}\,O_{1c}^{(i)}\,,
\end{equation}
where $a,b,c$ are adjoint indices, $T_a$ are Hermitian generators of $\mathfrak{g}$ and $O^{(0)}_{1a}=\rho_{\mathcal{R}}(T_a)$.
{Notice that a spanning set of such matrices for any $\mathfrak{g}$ is given by $\mathcal{O}^{(i)}_{1,b}=\big(\rho_{\mathcal{R}}(T^a)\frac{\partial}{\partial X^a}\big)^{d_i-1}\partial_{X^b}C_{d_i}(X)$, where $C_{d_i}$ are the independent Casimirs of $\mathfrak{g}$}. These operators are linearly dependent if some Dynkin labels vanish. 
In terms of the solutions to \eqref{equivarianceAppendix} we construct the tensors 
\begin{equation}
\label{gijdef}
\mathsf{g}_{ij}=\frac{k^{ab}\tr_{\mathcal{R}}( O_{1a}^{(i)}O_{1b}^{(j)}  )}{\tr_{\mathcal{R}}(1)}\,,
\qquad
\mu_{ijk}=\frac{{i}f^{abc}\,\tr_{\mathcal{R}}( O_{1a}^{(i)}O_{1b}^{(j)}O_{1c}^{(k)}  )}{\tr_{\mathcal{R}}(1)},
    \end{equation}
\begin{equation}
\label{eq:x_and_b}
\mathsf{b}_{ijk\ell}=k^{ac} \,k^{bd}\,\frac{\tr_{\mathcal{R}}(O_{1a}^{(i)}O_{1b}^{(j)}O_{1c}^{(k)}O_{1d}^{(\ell)})}{\tr_{\mathcal{R}}(1)}\,,
\qquad\mathsf{x}_{ijk\ell}={-}\frac{{2}}{h^{\vee}}f^{ab}_{\,\,\,\,\,\,e} \,f^{cde}\,\frac{\tr_{\mathcal{R}}(O_{1a}^{(i)}O_{1b}^{(j)}O_{1c}^{(k)}O_{1d}^{(\ell)})}{\tr_{\mathcal{R}}(1)}\,,
\end{equation}
where $k^{ab}$ is the {inverse-}Killing form, {normalised so that $f_{cd}^{\,\,\,\,e}f_{eg}^{\,\,\,\,b}k^{ca}k^{dg}=-h^\vee k^{ab}$, and $h^{\vee}$ is the dual Coxeter number}. The tensors $\mathsf{g}$ and $\mu$ naturally appear when computing two- and three-point functions, whereas the last two arise  from the four point function of scalar operators at tree-level and for one-loop corrections respectively. The factor of $i$ in the definition of $\mu_{ijk}$ guarantees it is real. Indices $i,j,...$ are lowered and raised with the tensor $\mathsf{g}_{ij}$ and its inverse $\mathsf{g}^{ij}$.

A variety of properties obeyed by these tensors can be derived by using the cyclicity of the trace, standard properties of the structure constants $f_{ab}^{\,\,\,c}$ and the equivariance condition \eqref{equivarianceAppendix}. For example, an elementary computation shows that $k^{ab}O^{(i)}_{1a}O^{(j)}_{1b}$ commutes with all $\rho_\mathcal{R}(T_a)$, therefore Schur's lemma implies
\begin{equation}
    k^{ab}O^{(i)}_{1a}O^{(j)}_{1b}=\mathsf{g}_{ij}\mathds{1}_\mathcal{R}\,.
\end{equation}
This identity is very useful when contracting adjacent exotic operators on the line. A second handy identity comes from noticing that the combination $f^{ab}_{\,\,\,\,c}O^{(i)}_{1a}O^{(j)}_{1b}$ also solves equation \eqref{equivarianceAppendix} and therefore can be expressed as a sum over $k$ of operators $O^{(k)}_{1c}$. Multiplying both sides of this expansion by $O_1$ with free indices and taking the trace proves the identity
\begin{equation}
    {i}f^{ab}_{\,\,\,\,\,\,c}\,O_{1a}^{(i)}O_{1b}^{(j)}=\mu_{ij}^{\,\,\,\,k}\,O_{1c}^{(k)}\,,
\end{equation}
hence we see that the tensor $\mu_{ijk}$ allows to define a multiplication on $\mathcal{M}_1$, which is known to be associative and commutative (see \cite{Okubo:1981}). Furthermore, inserting this equality twice in the definition of  $\mathsf{x}_{ijk\ell}$ we obtain
\begin{equation}
\label{cXandmumu}
h^{\vee}\,\mathsf{x}_{ijk\ell}={2}\mu_{ij}^{\,\,\,\,\,e}\,\mathsf{g}_{e e'}\,\mu_{k\ell}^{\,\,\,\,\,e'}\,.
\end{equation}
The symmetries $\mu_{ijk}=\mu_{jki}=\mu_{kij}$ under even permutations follow trivially from the cyclicity of the trace, but the highly non-obvious commutativity of the multiplication on $\mathcal{M}_1$  further implies complete symmetry $\mu_{ijk}=\mu_{(ijk)}$. Other identities that can be demonstrated using similar arguments are the symmetry of $\mathsf{x}_{ijkl}$ and $\mathsf{b}_{ijk\ell}-\mathsf{b}_{jik\ell}=\mathsf{g}_{ik}\mathsf{g}_{jl}-\mathsf{g}_{il}\mathsf{g}_{jk}$.

%%%%%%%%%%%%%%%%%%%%%%%%%%%%%%%%

At one-loop level, one should also consider corrections to the two-point and four-point functions. In the first case there are two inequivalent structures, depending on where the gluon attaches to the line: one is $\mu_{0ij}={-\frac{h^\vee}{2}\mathsf{g}_{ij}}$, the other
${\frac{i}{\tr_{\mathcal{R}}(1)}}f^{abc}\,\tr_{\mathcal{R}}( O_{1b}^{(i)}O_{1a}^{(0)} O_{1c}^{(j)}  )$, but they are both equal to ${\frac{h^{\vee}}{2}} \mathsf g_{ij}$ with opposite signs. Analogously, for the correction to the four-point function one has to consider structures of the type
\begin{equation}
    \mathsf{T}_{ijkl}^{(1,1)} \coloneq {\frac{i}{\tr_{\mathcal{R}}(1)}}f^{abc}  k^{de} \tr_\mathcal{R} (O_{1a}^{(0)} O_{1b}^{(i)} O_{1d}^{(j)} O_{1c}^{(k)} O_{1e}^{(l)} ),
\end{equation}
with the first of the two indices $(1,1)$ indicating that the gluon attaches to the first of the two propagators (the one from the first insertion to the third), the second that on the line it is the first insertion out of five. Another example is
\begin{equation}
    \mathsf{T}_{ijkl}^{(1,2)} \coloneq {\frac{i}{\tr_{\mathcal{R}}(1)}}f^{abc}  k^{de} \tr_\mathcal{R} ( O_{1b}^{(i)}O_{1a}^{(0)} O_{1d}^{(j)} O_{1c}^{(k)} O_{1e}^{(l)} ).
\end{equation} 
These are not independent colour structures as each of them is, up to a sign, equal to $\mathsf b_{ijkl}$, specifically $\mathsf T^{(1,1)}_{ijkl}=\mathsf T^{(1,4)}_{ijkl}=-\mathsf T^{(1,2)}_{ijkl}=-\mathsf T^{(1,3)}_{ijkl}= {-\frac{h^{\vee}}{2}} \mathsf{b}_{ijkl}$.

\vspace{0.3cm}

\noindent
{ \bf Specialization to $\mathfrak{g}=\mathfrak{su}(N)$.}
Here we show how to efficiently compute the tensors above for $\mathfrak{su}(N)$. Similar techniques exist for the other classical groups as well.
The relevant commutation relations are
\begin{equation}
[J^{A}_B,\,J^{C}_D]\,=\delta^{C}_B\,J^{A}_D-\,\delta^{A}_D\,J^{C}_B\,,
\end{equation}
where $A,B,..=1,\dots ,N$ and $J^A_A=0$ (sum over repeated indices is understood) {and we also assume the Hermiticity property $\big(J^A_B\big)^\dagger = J^B_A$.} In double index notation the Killing form and structure constants {(we don't include a factor of i in the definition, since we are working in a non-Hermitian basis)} take the form 
\begin{equation}
k^{AC}_{BD}={\frac{1}{2}}\Big(\delta^A_{{D}} \delta^{{C}}_{{B}}- \frac{1}{N}\delta^A_B \delta^C_D\Big)\,,
\qquad
F^{A_1 A_2 {B_3}}_{B_1 B_2 {A_3}} = (\delta^{A_1}_{{A_3}}\delta^{A_2}_{B_1}\delta^{{B_3}}_{B_2}-\delta^{A_1}_{B_2}\delta^{A_2}_{{A_3}}\delta^{{B_3}}_{B_1})\,.
\end{equation} 
A basis of solutions to the equation \eqref{equivarianceAppendix} is given by
\begin{equation}
    \big(O_1^{(i)}\big)^A_{B}=\big(J^{(i+1)}\big)^A_B\coloneq  J^{A}_{C_1} J^{C_1}_{C_2}\dots J^{C_{i}}_{B}
    \,,
\end{equation}
and notice these operator-valued matrices are Hermitian, as expected. In this basis, the two point function \eqref{gijdef} is easily computed to be
\begin{equation}
\label{metricandCasimirsSUN}
    \mathsf{g}_{ij} = {\frac{1}{2}}\big(C_{i+j+2}-\frac{1}{N}C_{i+1}C_{j+1}\big) \,,
    \qquad
\text{where}\,\,\,
\big(J^{(p)}\big)^A_A=C_p\,\mathds{1}_{\mathcal{R}}\,.
\end{equation}
It is useful to work in an orthogonal basis of operators by defining  
\begin{equation}
    O_{1a}^{(i)} \rightarrow  \widehat O_{1a}^{(i)}  \coloneq  O_{1a}^{(i)} + \sum_{j<i} K_{ij} O^{(j)}_{1a}
\end{equation}
in such a way that 
\begin{equation}
   \hat{\mathsf{g}}_{ij} = \delta_{ij}\mathsf h_{i}.
\end{equation}
This procedure gives $\mathsf{h}_{\ell} = {\frac{1}{2}\frac{\Delta_{\ell+1}}{\Delta_{\ell}}}$, where $\Delta_{\ell} \coloneq \det_{0 \leq \ell_1,\ell_2 \leq \ell} C_{\ell_1+\ell_2}$.
The Casimir $C_p$ defined in \eqref{metricandCasimirsSUN}  can be easily evaluated in terms of Dynkin labels as follows (see \cite{PP1968}):
given a partition $(\lambda_1, \lambda_2,\dots ,\lambda_{N-1},0)$, one 
considers the shifted weights $m_i = \lambda_i-\frac{\lambda}{N}$, where $\lambda=\sum_i \lambda_i$. Then take $\theta_{ij}$ to be the strictly upper triangular matrix with all non-zero entries equal to one, the Casimirs are then explicitly computed by first building the matrix
\begin{equation}
    a_{ij}=(m_i+N-i)\delta_{ij} - \theta_{ij}\,,
\end{equation}
and then summing elements of its $p$-th power
\begin{equation}
    C_p= \sum_{i,j=1}^N (a^p)_{ij}.
\end{equation}
This can be extended to all semisimple Lie groups with at least one nontrivial irreducible representation.

To compute the remaining tensor structures and rewrite them in terms of Casimirs $C_p$, it is convenient to 
collect the commutation relations of $(J^{(p)})^A_B$. This can be done efficiently using the evaluation homomorphism $\zeta:Y (\mathfrak{gl}_N) \rightarrow  \text{U}(\mathfrak{gl}_N)$, see \cite{Molev:1994}, where $Y (\mathfrak{gl}_N)$ denotes the Yangian.
In terms of the generators
  $t_{ij}^{(M)} \in Y (\mathfrak{gl}_N)$
  that satisfy the commutation relations
\begin{equation}
    [t_{ij}^{(M)},t_{kl}^{(L)}] = \sum_{r=0}^{\min(M,L)-1} t^{(r)}_{{kj}} t^{(M+L-r-1)}_{il}-t^{(M+L-r-1)}_{{kj}} t^{(r)}_{il}\,,
\end{equation}
the homomorphism acts as  $\zeta\big(t_{ij}^{(p)}\big) = \big(J^{(p)}\big)^i_j$.

\vspace{1cm}

%---------------
\section{III: Feynman integrals}
The main tool we use to compute correlation functions of defect operators in a perturbative expansion for small $g_{YM}$ is Feynman diagrams in position space. We report the results for each Feynman diagram contributing to $\vev{\mathcal O_1^{(i)}(t_0)\mathcal O_1^{(j)}(t_1)\mathcal O_1^{(k)}(t_2)\mathcal O_1^{(l)}(t_3)}$ at NLO. To regularize UV divergences we use point splitting regularization, where given two points $\vec{x}_1$ and $\vec{x}_2$ their distance is always taken to be $|\vec{x}_1-\vec{x}_2|>a$. Note that in all integrals we strip away the $R-$symmetry factor {and that in this section we work with operators that are not unit normalised $\mathcal O_{(i,I)}= O_{1a}^{(i)}\,\Phi^a_I$.}
\paragraph{Bulk diagrams} There are three kinds of diagrams involving vertices from the bulk Lagrangian: self-energy, four-scalar and gluon exchange diagrams. They can all be computed from the integrals and insertion rules presented in the modern notation in \cite{Beisert:2002bb,Drukker:2008pi}. The self energy diagrams in the normalization we choose are 
\begin{equation}
 \DefectSSSSSelfEnergyOne =  -2h^\vee g_{YM}^6\mathsf{b}_{ijkl}\frac{\log (t_2-t_0)-\log (a )+1}{{128} \pi ^6 (t_0-t_2)^2 (t_1-t_3)^2}  \,,
\end{equation}
and
\begin{equation}
 \DefectSSSSSelfEnergyTwo =   -2h^\vee g_{YM}^6\mathsf{b}_{ijkl}\frac{\log (t_3-t_1)-\log (a )+1}{{128} \pi ^6 (t_0-t_2)^2 (t_1-t_3)^2}\,.
\end{equation}
The four-scalar diagram is well known in the literature \cite{Usyukina:1994iw,Usyukina:1994eg} as it is the only one contributing to this correlator for the fundamental representation at large $N$ and it reads
\begin{equation}
 \DefectSSSSX = -\frac{g_{YM}^6 h^\vee \mathsf{x}_{ijkl}}{{128} \pi ^6 (t_0-t_2) (t_1-t_3)} \left(\frac{\log \left(\frac{(t_1-t_2) (t_0-t_3)}{(t_0-t_2) (t_1-t_3)}\right)}{(t_0-t_1) (t_2-t_3)} + \frac{\log \left(\frac{(t_0-t_1)
   (t_2-t_3)}{(t_0-t_2) (t_1-t_3)}\right)}{(t_1-t_2) (t_0-t_3)}\right)   \,.
\end{equation}
Finally, the gluon exchange diagram can also be computed starting from the integrals in \cite{Beisert:2002bb,Drukker:2008pi}; however, its prefactor coming from Wick contractions is null for any representation and any set of dimension-one operators
\begin{equation}
 \DefectSSSSH =  0  \,.
\end{equation}
\paragraph{Defect diagrams} Beyond the diagrams only generated by bulk interactions, there are two diagrams (each made of five intervals) that also involve an integral along the line. These diagrams are 
\begin{equation}
    \DefectSSSSYOne = -h^\vee  g_{YM}^6\mathsf{b}_{ijkl}\frac{ \left(6 \log (a)-6 \log (t_2-t_0)+\pi ^2-6\right)}{{384} \pi ^6 (t_0-t_2)^2 (t_1-t_3)^2} \,,
\end{equation}
and 
\begin{equation}
    \DefectSSSSYTwo = -h^\vee  g_{YM}^6\mathsf{b}_{ijkl}\frac{ \left(6 \log (a)-6 \log (t_3-t_1)+\pi ^2-6\right)}{{384} \pi ^6 (t_0-t_2)^2 (t_1-t_3)^2} \,.
\end{equation}
Summing all the diagrams and replacing the free coefficients with their expressions in terms of tensors in four indices and accounting for normalization, we obtain Eq.~\eqref{F1andf1}.\\
\paragraph{The two point function at NLO} Here we include the integrals necessary to compute the NLO two point function between operators $\mathcal O_1^{(i)}$. The first diagram we introduce is the self-energy
\begin{equation}
    \DefectSelfEnergy = \frac{{g_{YM}^4} h^\vee (-\log (t_2-t_1)+\log (a )-1) \mathsf{g}_{ij}}{{16} \pi ^4 (t_1-t_2)^2}\,.
\end{equation}
The second is the sum of three integrals -- similar to the defect diagrams above -- and it is
\begin{equation}
\DefectY     = -\frac{{g_{YM}^4}h^\vee \left(-6 \log (t_2-t_1)+6 \log (a )+\pi ^2-6\right) \mathsf{g}_{ij}}{{96} \pi ^4 (t_1-t_2)^2}.
\end{equation}
Together, they represent the NLO contribution to the two-point function of two dimension-one operators. We highlight that the sum of the diagrams is finite and that its tensor-structure is the same as the LO contribution. This means that the orthogonality condition $\hat{\mathsf{g}}_{ij} = \mathsf{h}_i\delta_{ij}$ does not get a correction from NLO diagrams. These diagrams are the ones we use to fix the topological sector and the integration constant in the perturbative bootstrap of $f_{ijkl}^{(1)}(\chi)$.
\section{IV: Integrating the four-point function non-perturbatively}
\label{app:integrating_f}
Let the four-point function of four dimension-one operators be defined as in Eq.~\eqref{eq:4pt_both_formulas}, using the convention where the correlator can be expressed as a prefactor multiplying a total derivative and the constant $\mathbb{F}_{ijkl}$. We can state that Eq.~\eqref{eq:4pt_both_formulas} is sufficient to compute the integrated correlator of the four-point function non-perturbatively. To prove this fact, we show that the integrand including the prefactor can be written as a total derivative in an appropriate variable. Considering without loss of generality the interval $t_0<t_1<t_2<t_3$ we can write
\begin{equation}
   \int_{-\infty}^{t_1} \vev{D_1^{(i)}(t_0)D_1^{(j)}(t_1)D_1^{(k)}(t_2)D_1^{(l)}(t_3)} dt_0 = \int_{\chi\left(-\infty\right)}^{\chi\left(t_1\right)} \left(\frac{dF(\chi)}{d\chi} \right) d\chi\,
\end{equation}
where $F(\chi)$ can be written in terms of OPE coefficients. The four point function in \eqref{eq:4pt_both_formulas} consists of a prefactor multiplying a constant and the derivative of a function of the cross ratio $\chi$. By changing integration variable using
\begin{equation}
    t_0 = \frac{t_2 t_3 \chi -t_1 (t_2 (\chi -1)+t_3)}{t_1 (-\chi )+t_3 (\chi -1)+t_2}
\end{equation}
we obtain that the kinematic prefactor multiplying the total derivative is
\begin{equation}
    (02)(13) \frac{dt_0}{d\chi} = \frac{1}{t_{12}t_{13}t_{23}}\,,
\end{equation}
which does not depend on the integration variable $\chi$. The integral then can be performed without knowing the full function $f_{ijkl}(\chi)$ and evaluated by considering the value of the primitive function at the end point of the integration domain $t_0\rightarrow t_1$; in this limit, $f_{ijkl}(\chi)$ can be expanded in conformal blocks. It is sufficient to restrict ourselves to the exchange of the identity, $D_2$ and $B_1[2,0]$ operators as all other contributions vanish when taking $t_0\rightarrow t_1$ (corresponding to $\chi \rightarrow 0$). From \cite{Liendo:2018ukf} we see that in this limit
\begin{equation}
    f_{ijkl}(\chi) =  \mathsf{g}_{ij} \mathsf{g}_{kl} \chi {+}\frac{1}{2} \chi^2 \left( \mathsf{g}_{ij} \mathsf{g}_{kl}-\mathbb{F}_{ijkl}-2 (C^2_{B_1[2,0]})_{ijk\ell}\right)+o\left(\chi^2\right)
\end{equation}
%and where the $o\left(\chi^2\right)\sim \chi^{1+\Delta}$ with $\Delta>1$ is associated to the lightest long operator.
and  the integral can finally be expressed only in terms of OPE coefficients. The minus sign next to $(C^2_{B_1[2,0]})_{ijk\ell}$ arises from the conventions of conformal blocks: near the unitarity bound we find $f_{B_1[2,0]}(\chi)=\lim_{\Delta\to 1}(\Delta-1)f_{\mathcal{L}^\Delta_{[0,0]}}(\chi)=-\chi^2+O(\chi^3)$. It is important to observe that the weak-coupling expansion is not uniform in the small-$\chi$ limit. In particular, the limit $\chi\to 0$ must be taken at fixed nonzero $g_{\mathrm{YM}}$, before expanding in the coupling. Reversing the order of limits generates terms such as $\chi^2\log\chi$, which are not subleading with respect to $\chi^2$ and invalidate the expansion above. The same mechanism works for the other three integration intervals, where we use Eq.~\eqref{eq:crossing} whenever the function $f_{ijkl}(\chi)$ must be expanded around $\chi\rightarrow 1$. We highlight that we can follow the same argument presented here to compute the integrated four-point function $\vev{D_1^{(i)}(t_0)D_1^{(j)}(t_1)D_n^{(k)}(t_2)D_n^{(l)}(t_3)} $ explicitly in terms of OPE coefficients, from which we can extract constraints similar to Eq.~\eqref{eq:constrain_for_displacement} by specializing one of the dimension-one operators to the displacement.

\section{V: {Infrared stability of half-BPS lines}}
Near the half-BPS line we have a $5\mathfrak{m}_{\mathcal{R}}$-dimensional space of deformations parametrised by $\zeta^{(i,I)}$ and in the main text we have shown that quantum effects induce an RG flow and conformality is generically broken. Denote the exotic directions orthogonal to the super-displacement by $\hat{\mathcal{I}} = (\hat{i},I)$, $\hat{i}=1,...,\mathfrak{m}_{\mathcal{R}}-1$, then we wish to show that flowing towards the IR causes a generic initial point to approach the BPS-locus. To aid with this task, we define the exotic separation of a generic point from the BPS-subspace as $R_{ex}^2 = \mathsf{g}_{\hat{i}\hat{j}}\zeta^{\hat{i}}\cdot \zeta^{\hat{j}}$, from where we find
\begin{equation}
    \frac{d R_{ex}^2}{d\log(1/a)} = 2\mathsf{g}_{\hat{i}\hat{j}}\zeta^{\hat{i}}\cdot\beta^{\hat{j}}\,.
\end{equation}
Showing that this quantity is positive is equivalent to demonstrating that the BPS-locus is IR attractive. If we then use \eqref{betafunctionwithC} in combination with the selection rule 
\eqref{eq:constrain_for_displacement}, we arrive at the equation
\begin{equation}
    \mathsf{g}_{\hat{i}\hat{j}}\zeta^{\hat{i}}\cdot \beta^{\hat{j}} = 2(\zeta^{\hat{i}}\cdot \zeta^{\hat{j}})\big(C^2_{B_1[2,0]^+}\big)_{\hat{i}\hat{j},\hat{k}\hat{l}}(\zeta^{\hat{k}}\cdot \zeta^{\hat{l}})\,+\,O(\zeta^6)\,.
\end{equation}
Next we go to a basis of $B_1[2,0]^+$ super-multiplets that is orthonormal and has a metric $G_{AB}=\delta_{AB}$, with $A,B =1,...,\frac{\mathfrak{m}_{\mathcal{R}}(\mathfrak{m}_{\mathcal{R}}-1)}{2}$. In this case we may write
\begin{equation}\label{zeta_beta_eqn_positive}
    \mathsf{g}_{\hat{i}\hat{j}}\zeta^{\hat{i}}\cdot \beta^{\hat{j}} = 2 \sum_{A}\Big[\sum_{\hat{i},\hat{j}}(\zeta^{\hat{i}}\cdot \zeta^{\hat{j}})C_{\hat{i}\hat{j}}^A\Big]^2
    \,+\,O(\zeta^6)\,,
\end{equation}
which is manifestly non-negative, since OPE coefficients can be chosen to be real in a basis where the two-point functions of parity-even operators are positive definite.
Finally, notice that a basis for $B_1[2,0]^+$ operators consists of $\mathcal{O}^{(\hat{i},\hat{j})}+\mathcal{O}^{(\hat{j},\hat{i})}+\text{$\mathfrak{Q}$-exact}$, where $\mathcal{O}^{(i,j)}$ is defined in  Eq.~\eqref{eq:B1[2,0]} and the $\mathfrak{Q}$-exact terms are linear combinations of $\mathcal{O}^{(0,k)}+\mathcal{O}^{(k,0)}$, $k=0,\dots,\Monedim-1$. Hence we can view the coefficients $C_{\hat{i}\hat{j}}^A$ as a change of basis matrix between the orthonormal and symmetric-pair bases. For half-BPS Wilson lines this matrix is non-degenerate. A degeneracy would correspond to an extra relation in the quarter-BPS chiral ring. There is no such relation in our case, as we showed by turning on the Yang-Mills coupling. This proves the positivity of \eqref{zeta_beta_eqn_positive}.

One may wonder if a stronger condition holds, namely that
%\col{A necessary and sufficient condition for the eigenvalues of the reduced anomalous dimension matrix to be positive is that }
\begin{equation}
\label{Npositive}
\mathcal{N}:=\frac{1}{2}v^{\hat{\mathcal{I}}}
\, g_{\mathcal{\hat{I}}\mathcal{\hat{K}}}
\frac{\partial \beta^{\hat{\mathcal{K}}}}{\partial \zeta^{\hat{\mathcal{J}}}}
v^{\hat{\mathcal{J}}}>0\,,
\end{equation}
for every nonzero real vector \(v\) and every sufficiently small nonzero exotic deformation \(\zeta\).
For situations in which the  beta-function starts with a linear term in $\zeta$, positivity of the corresponding quadratic form is equivalent to the decrease of the squared radius $g\zeta \zeta$ towards the infrared in every direction.
For a flow beginning at cubic order, which is often the case in supersymmetric theories, see e.g.~\cite{Green:2010da}  strict radial attraction is instead a weaker condition, already sufficient for local infrared stability. The stronger condition additionally ensures that infinitesimally neighbouring trajectories approach one another at equal RG scale, as measured with $g$, for as long as they remain in this neighbourhood

In our case we can show that, at least in the vicinity of $g_{YM}=0$, the stronger condition holds. This follows from a curious property of the squared OPE coefficient tensor $C^{\,2}_{B_1[2,0]^{+}}$, namely that it is totally symmetric in its four indices.
Or, in other words, that it satisfies a crossing equation by itself.
We verify the total symmetry property at leading order in  $g^2_{YM}$. The explicit form of this tensor in this limit is obtained by inserting \eqref{F0andf0} in  \eqref{C2andL} and symmetrizing in $i,j$ to single out the exchange of parity even operators and is given by
\begin{equation}
\label{C0ijklAPP}
2\bigl(C^{\,2}_{B_1[2,0]^{+}}\bigr)_{ij,k\ell}^{(0)}=\mathsf{g}_{ij} \mathsf{g}_{k\ell}+\mathsf{g}_{ik} \mathsf{g}_{j\ell}+\mathsf{g}_{i\ell} \mathsf{g}_{jk}-\mathbb{F}^{(0)}_{ijk\ell}
-\mathsf{x}_{ijk\ell}\,.
\end{equation}
This expression makes total symmetry manifest.
%Above we restricted the sum to indices $\hat{\mathcal{I}},\hat{\mathcal{J}},\dots $ which are orthogonal to the super-displacement.
%
A simple calculation, using \eqref{eq:constrain_for_displacement}, \eqref{betafunctionwithC},  shows that
\begin{equation}
\label{Nstep1}
\mathcal{N}=\left((v^{\hat{i}}\cdot v^{\hat{j}})(\zeta^{\hat{k}}\cdot \zeta^{\hat{\ell}})+2 (v^{\hat{i}}\cdot \zeta^{\hat{j}})(v^{\hat{k}}\cdot \zeta^{\hat{\ell}})\right)
\bigl(C^{\,2}_{B_1[2,0]^{+}}\bigr)_{\hat{i}\hat{j},\hat{k}\hat{\ell}}\,+\,O(\zeta^4).
\end{equation}
If  $C^{\,2}_{B_1[2,0]^{+}}$ is totally symmetric, we can write it  as $\sum_A C^{A}_{\hat{i}\hat{k}} C^{A}_{\hat{j}\hat{\ell}}$ and as $\sum_A C^{A}_{\hat{i}\hat{j}} C^{A}_{\hat{k}\hat{\ell}}$
for the first and  second term in \eqref{Nstep1} respectively.
We recognize that, up to higher orders in $\zeta$, this is a sum of positive terms  
\begin{equation}
\mathcal{N} = \sum_{A,I,J}
\Big(\sum_{\hat{i},\hat{j}} C^{A}_{\hat{i}\hat{j}}\,v^{\hat{i} I}\zeta^{\hat{j} J}\Big)^2 + 2\sum_A
\Big(\sum_{\hat{i},\hat{j}} C^{A}_{\hat{i}\hat{j}}\,v^{\hat{i}}\cdot\zeta^{\hat{j}}
\Big)^2\,+\,O(\zeta^4)\,.
\end{equation}
This shows that total symmetry of $C^{\,2}_{B_1[2,0]^{+}}$, together with the non-degeneracy of $C^{A}_{\hat{i}\hat{j}}$, implies the stronger condition.
Should $C^{\,2}_{B_1[2,0]^{+}}$ fail to be totally symmetric at higher orders in $g_{YM}$, the strong condition would still hold in a neighbourhood of $g_{YM}=0$.
\section{VI: Extracting anomalous dimensions}

The anomalous dimensions of the operators whose super-primary at $g^2_{YM}=0$ has quantum numbers $\Delta_0=1$,  $[a,b]=[0,0]$, $s=0$ can be computed using a trick that relies on the mechanism of multiplet recombination, similar to \cite{Anselmi:1998ms,Belitsky:2007jp,Rychkov:2015naa}. The result is that these operators get all the same anomalous dimension at first order 
\begin{equation}
\label{AndimAppendixResult}
\Delta=1+\hat{\lambda}_{\mathfrak{g}}+\dots\,,
\qquad
\hat{\lambda}_{\mathfrak{g}}=\tfrac{g^2_{YM}h_{\mathfrak{g}}^{\vee}}{4\pi^2}\,.
\end{equation}
 For $\mathfrak{g}=\mathfrak{su}(N)$, $h^{\vee}=N$ and this coincides with the result first obtained in \cite{Alday:2007he}  for the fundamental Wilson line.

We now present the derivation.
On the one hand one can compute the norm of the supersymmetric descendant that becomes null when $\Delta\rightarrow 1$. Denote by $|\Delta\rangle$ the super-primary state and the relevant descendant by
\begin{equation}
|X(v)\rangle=\tfrac{1}{2} v^A v^B\epsilon^{\alpha\beta}\mathfrak{Q}_{\alpha A}\mathfrak{Q}_{\beta B}|\Delta\rangle\,,
\end{equation}
where $A,B=1,\dots, 4$ are $\mathfrak{sp}(4)$ fundamental indices, see \cite{Ferrero:2023znz} for our conventions. A simple standard computation reveals that  
\begin{equation}
\label{gammaformulaappendix}
\langle X(v_1)|X(v_2)\rangle=\,\Delta(\Delta-1)(\bar{v}_1 v_2)^2
\langle \Delta|\Delta\rangle\,,
\,\,\,\,
\xRightarrow{\text{$\Delta=1+\gamma+\dots$}}
\,\,\,\,
\gamma=
\frac{\langle X(v_1)|X(v_2)\rangle^{(0)}}{(\bar{v}_1 v_2)^2\langle \Delta|\Delta\rangle^{(0)}}\,,
\end{equation}
where $\langle \dots \rangle^{(0)}$ means that it is computed at leading order in $g^2_{YM}$.
In gauge theory the operators at $
\Delta_0=1$ are $O_1^{(i)}(\Phi_6)$. Their supersymmetric variations are determined by
\begin{equation}
\mathfrak{Q}_{\alpha A}\Phi^6=\Psi_{\alpha A}\,,
\qquad 
\tfrac{1}{2}v^A v^B\epsilon^{\alpha \beta}\mathfrak{Q}_{\beta B}\Psi_{\alpha A}=\,
\tfrac{1}{4}\,
(v\,\Gamma_{IJ}\,v)[\Phi^I,\Phi^J ]\,,
\end{equation}
where $\Gamma_{IJ}=\tfrac{1}{2}[\Gamma_I,\Gamma_J]$ with $\Gamma_{I}$ $SO(5)$ gamma matrices; here they serve to implement the equivalence between the symmetric product of two spinor representations and the antisymmetric product of two vector representations.
In these conventions, see \cite{Cooke:2017qgm} for more details,  the scalar propagator is given by $\frac{g^2_{YM}}{{4} \pi^2} \frac{k^{ab}}{x^2}\delta^{IJ}$, where $a,b$ are adjoint indices.

%The numerical factor $\sharp$ (which turns out to be $\tfrac{1}{4}$) can be fixed by looking at the standard SUSY transformations preserved by the line, see for example the conventions in  \cite{Cooke:2017qgm}. Here,  we will determine $\sharp$ by comparison with the known result for the fundamental Wilson line for $\mathfrak{su}(N)$. \DA{I am not a fan of leaving it as open coefficient in the final version. Maybe we could set it to 1/4 and highlight that this is also coherent with the fundamental line.}

Now we are ready to compute the relevant two point functions. For the primaries one finds
\begin{equation}
\langle 
O_1^{(i)}(\Phi_6(t_1))O_1^{(j)}(\Phi_6(t_2))\rangle^{(0)}=\,
\frac{g^2_{YM}}{{4} \pi^2}\frac{\mathsf{g}_{ij}}{t_{12}^2}
\,,
\end{equation}
where $\mathsf{g}_{ij}$ is given in \eqref{gijdef}.
For the descendants $X^{(i)}(t,v)=\tfrac{1}{4}(v\,\Gamma_{IJ}\,v)\, O_1^{(i)}([\Phi^I(t),\Phi^J(t)])$ we obtain
\begin{equation}
\langle 
X^{(i)}(t_1,v_1)
X^{(j)}(t_2,v_2)\rangle^{(0)}=\,
\left(\frac{g^2_{YM}}{{4} \pi^2}\right)^2\frac{h^{\vee}\mathsf{g}_{ij}}{t_{12}^4}(v_1\Omega v_2)^2\,,
\end{equation}
here we have two propagators, and two Wick contractions, instead of one, and we use the identity
\begin{equation}
(v_1\Gamma_{IJ}v_1)
(v_2\Gamma_{KL}v_2)\delta^{IK}\delta^{JL}=8(v_1\Omega v_2)^2\,.
\end{equation}
The dual Coxeter number appears using $f_{cd}^{e} f_{eg}^b k^{ca} k^{dg} ={-}h^{\vee} k^{ab}$.
The fact that all two point functions here are proportional to the same $\mathsf{g}_{ij}$ with the same proportionality constant implies that the degeneracy is not lifted at first order. From  \eqref{gammaformulaappendix} we extract the anomalous dimension
\begin{equation}
\gamma=\frac{g^2_{YM}h^{\vee}}{4\pi^2}\,,
\end{equation}
which reduces to the result \eqref{AndimAppendixResult}.

We can apply a similar logic to the three-point function. So we compute
\begin{equation}
\langle O_1^{(i)}(X(t_1))O_1^{(j)}(Y(t_2))O_1^{(k)}([\bar{X},\bar{Y}](t_3))\rangle^{(0)}=
\left(\frac{g^2_{YM}}{{4} \pi^2}\right)^2\,\frac{\mu_{ijk}}{t^2_{13}t_{23}^2}\,
\end{equation}
and 
\begin{equation}
\langle O_1^{(i)}([X,Y](t_1))O_1^{(j)}([\bar{X},\bar{Y}](t_2))\rangle^{(0)}=h^{\vee}
\left(\frac{g^2_{YM}}{{4} \pi^2}\right)^2\,
\frac{\mathsf{g}_{ij}}{t^4_{12}}\,.
\end{equation}
This implies that the exchange of the $B_1[2,0]$ operators in the free theory  that are recombined is $\frac{1}{h^{\vee}}\mu_{ijs}\mu_{k\ell}^{\,\,\,\,\,s}={\frac12}\mathsf{x}_{ijk\ell}$ (the factor of $\left(\frac{g^2_{YM}}{{4}\pi^2}\right)^4$ is cancelled by the normalization of $D_1^{(i)}$).

\section{VII: Deforming the ordinary Wilson line}
\label{app:ordinary_line}
In this appendix we derive the RG flow of Fig.~\ref{fig:rg_flow} with the conformal perturbation theory of the main text, now applied to the ordinary Wilson line. 
%We use the Killing form of \eqref{metricandCasimirsSUN}, i.e.\ $\tr_{\square}(T^aT^b)=\delta^{ab}$ and $k^{ab}=\delta^{ab}$, for which $f_{cd}^{\,\,\,\,e}f_{eg}^{\,\,\,\,b}k^{ca}k^{dg}=2h^\vee k^{ab}$. 
We normalise the scalar propagator as
$
\langle\Phi^a_M(x)\Phi^b_N(0)\rangle=\frac{g^2_{YM}}{{4}\pi^2}\,\frac{k^{ab}}{x^2}\,\delta_{MN}\,,
$
and consider the generalised Wilson line
\begin{equation}
\label{Wxi}
\mathcal{W}_\xi=\mathds{P}\exp\int dt\,\Big(iA_t+\xi^{\mathcal{M}}\,\mathcal{O}_{\mathcal{M}}\Big)\,,\qquad 
\mathcal{O}_{\mathcal{M}}=O^{(i)}_{1a}\,\Phi^a_M\,,\qquad \mathcal{M}=(i,M)\,,
\end{equation}
with $i=0,\dots,\Monedim-1$ and $M=1,\dots,6$. The ordinary line is $\xi=0$ and the half-BPS line is $\xi^{(i,M)}=\delta^{i0}n^M$, with $|n|=1$. In these conventions, the operators $\mathcal{O}_{\mathcal{M}}$ are not unit normalised:
$
G_{\mathcal{M}\mathcal{N}}=\langle \mathcal{O}_{\mathcal{M}}(0)\mathcal{O}_{\mathcal{N}}(\infty)\rangle={\frac{\hat{\lambda}_{\mathfrak{g}}}{h^\vee}\,\mathsf{g}_{ij}\,\delta_{MN}}\,.
$

For relevant deformations, conformal perturbation theory \cite{Komargodski:2016auf,Behan:2017mwi} gives
\begin{equation}
\label{betaordinary}
\beta^{\mathcal{M}}=(\Delta_{\mathcal{M}}-1)\,\xi^{\mathcal{M}}
-\frac{1}{3}\,\xi^{\mathcal{N}} \xi^{\mathcal{K}} \xi^{\mathcal{L}}\,
\delta_{\mathcal{N}}C_{\mathcal{K}\mathcal{L}}^{\,\,\,\,\,\,\,\,\mathcal{M}}+\dots\,,
\end{equation}
where $\delta_{\mathcal{M}}C_{\mathcal{N}\mathcal{K}\mathcal{L}}$ is defined as in \eqref{C_IJKL_int_formula}, now on the ordinary line, and indices are raised with $G^{\mathcal{M}\mathcal{N}}$. The quadratic term in $\xi$ is absent because the ordinary Wilson line DCFT has an  $SO(6)$ R-symmetry. As we will shortly review the operators $\mathcal{O}_{\mathcal{M}}$ of this DCFT are marginal for $g_{YM}=0$ but acquire a negative anomalous dimension at one loop.

%Each propagator carries a factor $\hat{\lambda}_{\mathfrak{g}}/2h^\vee$. Hence $\delta_{\mathcal{M}}C_{\mathcal{N}\mathcal{K}\mathcal{L}}$, obtained from the free four-point function, is $O(\hat{\lambda}_{\mathfrak{g}}^2)$, and raising one index leaves $O(\hat{\lambda}_{\mathfrak{g}})$. 
Power counting arguments show that the linear and cubic terms are of the same order after raising one index, while the term of order $\xi^{2n+1}$, coming from free $(2n+2)$-point functions, is $O(\hat{\lambda}_{\mathfrak{g}}^{\,n})$. At one loop the beta-function is thus exactly cubic and reliable for $\xi=O(1)$, as in \cite{Polchinski:2011im}. 
The analysis does not require $|\xi|\ll1$, but only $|\xi|\ll 1/g_{YM}$.

\paragraph{Linear term.} The one-loop anomalous dimension is the same for all $\mathcal{O}_{\mathcal{M}}$:
\begin{equation}
\Delta_{\mathcal{M}}=1-\frac{\hat{\lambda}_{\mathfrak{g}}}{2}+\dots\,.
\end{equation}
It comes from the gluon exchanged between two points of the line on opposite sides of the insertion, which on the half-BPS line is cancelled by the $\Phi^6$ exchange. 
%By \eqref{equivarianceAppendix}, its colour factor relative to $\langle\tr_{\mathcal{R}}\mathcal{W}_0\rangle$ is $-h^\vee\mathsf{g}_{ij}\delta_{MN}$ for every $O^{(i)}_{1a}$, so the $\mathcal{O}_{\mathcal{M}}$ do not mix. 
For the fundamental representation this reproduces \cite{Alday:2007he}.

\paragraph{Cubic term.} At leading order the four-point function is given by Wick contractions. For ${t_1<t_2<t_3<t_4}$,
\begin{equation}
\langle \mathcal{O}_{(i,M)}\mathcal{O}_{(j,N)}\mathcal{O}_{(k,P)}\mathcal{O}_{(\ell,Q)}\rangle
=\Big(\frac{\hat{\lambda}_{\mathfrak{g}}}{h^\vee}\Big)^2\left[\frac{\delta_{MN}\delta_{PQ}\,\mathsf{g}_{ij}\mathsf{g}_{k\ell}}{t_{12}^2t_{34}^2}
+\frac{\delta_{MQ}\delta_{NP}\,\mathsf{g}_{i\ell}\mathsf{g}_{jk}}{t_{14}^2t_{23}^2}
+\frac{\delta_{MP}\delta_{NQ}\,\mathsf{b}_{ijk\ell}}{t_{13}^2t_{24}^2}\right],
\end{equation}
with the colour structures of \eqref{F0andf0}. When the integrated operator in \eqref{C_IJKL_int_formula} crosses another insertion, the cyclic order along the line changes, and so does the tensor multiplying a given Wick contraction. For instance, the contraction $(\mathcal{M}\mathcal{N})(\mathcal{K}\mathcal{L})$ comes with $\mathsf{g}_{ij}\mathsf{g}_{k\ell}$ for $t\notin(t_2,t_3)$ and with $\mathsf{b}_{jki\ell}$ for $t_2<t<t_3$, so that, up to the overall prefactor,
\begin{equation}
\int_{|t-t_1|>a} \frac{dt\,(\dots)}{(t-t_1)^2\,t_{32}^2}=\frac{2\,\mathsf{g}_{ij}\mathsf{g}_{k\ell}}{a\,t_{32}^2}+\frac{\mathsf{b}_{jki\ell}-\mathsf{g}_{ij}\mathsf{g}_{k\ell}}{t_{21}t_{31}t_{32}}\,.
\end{equation}
The first term is the exchange of the identity and is removed by the counterterm. %\footnote{The regulator must be imposed on $|t-t_i|$, consistently with the identity counterterm. A cutoff on the cross-ratio shifts the finite part by terms proportional to $\mathsf{g}\,\mathsf{g}$.}.
Treating the other two contractions in the same way, and using $\mathsf{b}_{ijk\ell}-\mathsf{b}_{jik\ell}=\mathsf{g}_{ik}\mathsf{g}_{j\ell}-\mathsf{g}_{i\ell}\mathsf{g}_{jk}$, we find
\begin{equation}
\delta_{\mathcal{M}}C_{\mathcal{N}\mathcal{P}\mathcal{Q}}=\Big(\frac{\hat{\lambda}_{\mathfrak{g}}}{h^\vee}\Big)^2\,\mathsf{S}_{ijk\ell}\,
\big(\delta_{MN}\delta_{PQ}+\delta_{MP}\delta_{NQ}+\delta_{MQ}\delta_{NP}\big)\,,
\qquad
\mathsf{S}_{ijk\ell}:=\mathsf{b}_{ijk\ell}-\mathsf{g}_{ik}\mathsf{g}_{j\ell}\,,
\end{equation}
where $\mathcal{P}=(k,P)$, $\mathcal{Q}=(\ell,Q)$ and $\mathsf{S}$ is totally symmetric.

\paragraph{Result.} Inserting both terms in \eqref{betaordinary},
\begin{equation}
\label{betaordinaryfinal}
\beta^{(i,M)}=-\frac{\hat{\lambda}_{\mathfrak{g}}}{2}\Big[\xi^{(i,M)}
+\frac{{2}}{h^\vee}\,\xi^{(\ell,M)}\,(\xi^{j}\cdot\xi^{k})\,\mathsf{S}_{jk\ell i'}\,\mathsf{g}^{i'i}\Big]+O(\hat{\lambda}_{\mathfrak{g}}^2)\,,
\end{equation}
where $\cdot$ denotes the $SO(6)$ scalar product. 
If we turn on only the canonical direction $i=0$ by setting $\xi^{(i,M)}=\delta_{i,0} \xi^{(0,M)}$ we recover the standard result as follows. Using $\mathsf{b}_{0000}-\mathsf{g}_{00}^2=-\frac{h^\vee}{{2}}\mathsf{g}_{00}$, we obtain $\beta^{(0,M)}=-\tfrac{\hat{\lambda}_{\mathfrak{g}}}{2}\,\xi^{(0,M)}\big(1-|\xi^{0}|^2\big)$ for any $\mathcal{R}$, reproducing the flow of \cite{Polchinski:2011im} for the fundamental representation and of \cite{Beccaria:2022bcr} for the symmetric one, with the half-BPS line at $|\xi^0|=1$.

\paragraph{Example: the adjoint of $\mathfrak{su}(3)$.} Here $\Monedim=2$ and realising the adjoint on traceless $3\times 3$ matrices $M$, we take $O^{(0)}_1(X)M=[X,M]$ and $O^{(1)}_1(X)M=\{X,M\}-\tfrac{2}{3}\tr_{\square}(XM)\,\mathds{1}$. {With the inner product $\langle M_1,M_2\rangle=\tr_\Box(M_1^\dagger M_2)$, both operators are Hermitian}. Setting $\mathbf{x}^M=\xi^{(0,M)}$ and $\mathbf{y}^M=\xi^{(1,M)}$, \eqref{betaordinaryfinal} becomes
\begin{equation}
\begin{aligned}
\beta^{\text{1-loop}}_{\mathbf{x}}&=-\tfrac{\hat{\lambda}_{\mathfrak{g}}}{2}\Big[\mathbf{x}\big(1-\mathbf{x}^2-\tfrac{5}{9}\mathbf{y}^2\big)-\tfrac{10}{9}(\mathbf{x}\cdot \mathbf{y})\,\mathbf{y}\Big]\,,\\
\beta^{\text{1-loop}}_{\mathbf{y}}&=-\tfrac{\hat{\lambda}_{\mathfrak{g}}}{2}\Big[\mathbf{y}\big(1-\mathbf{x}^2-\tfrac{13}{9}\mathbf{y}^2\big)-2(\mathbf{x}\cdot \mathbf{y})\,\mathbf{x}\Big]\,.
\end{aligned}
\end{equation}
It is not hard to see that for all the zeroes of these beta-functions the vectors $\mathbf{x}$ and $\mathbf{y}$ are parallel. We call this direction $n^{M}$, with $n^2=1$. In Fig.~\ref{fig:rg_flow} we plot the flow lines restricted to the two dimensional subspace of the twelve dimensional $(\mathbf{x},\mathbf{y})$ space labelled by $(x,y)$ with  $\mathbf{x}^M=x\,n^{M}$, $\mathbf{y}^M=y\,n^{M}$. Without loss of generality we take $n^M=\delta^{M6}$. 
Notice that at all the fixed points, with the exception of the ordinary line ($\xi=0$) the $SO(6)$ R-symmetry is broken to $SO(5)$ under which the longitudinal directions transform as a singlet while the transverse directions as five-vectors. 
%Fig.~\ref{fig:rg_flow} shows the $SO(5)$-invariant plane $x^M=x\,\delta^{M6}$, $y^M=y\,\delta^{M6}$.

In the following table we list the fixed point together with the eigenvalues
of the one-loop beta-function Jacobian for the $SU(3)$ adjoint line, in units of
$\widehat{\lambda}_{\mathfrak{su}(3)}
=3g_{YM}^{2}/(4\pi^{2})$.
Each transverse eigenvalue has multiplicity five.
The last column counts relevant directions at linear order,
corresponding to negative eigenvalues in the convention
$\beta=\mu\,d\xi/d\mu$.
The signs of the mixed fixed-point coordinates are independent.
%--------------------------------------------------------
\begin{center}
\renewcommand{\arraystretch}{1.2}
\setlength{\arraycolsep}{8pt}

% Dotted separators: no additional packages required.
\newcommand{\rgdots}{\ensuremath{\vcenter to 20pt{%
  \offinterlineskip
  \cleaders\vbox to 2.5pt{%
    \vfil\hbox{\rule{0.4pt}{0.4pt}}\vfil
  }\vfil
}}}

\(
\begin{array}{
  |l
  @{\hspace{6pt}\rgdots\hspace{6pt}}c
  @{\hspace{6pt}\rgdots\hspace{6pt}}c
  @{\hspace{6pt}\rgdots\hspace{6pt}}c
  @{\hspace{6pt}\rgdots\hspace{6pt}}c|
}
\hline
\mbox{Line}
& (x,y)
& \mbox{Longitudinal}
& \mbox{Transverse}
& n_{\mathrm{rel}}
\\
\hline
\mbox{Ordinary}
& (0,0)
& -\frac12,\;-\frac12
& -\frac12,\;-\frac12
& 12
\\
\mbox{Half-BPS}
& (\pm1,0)
& 1,\;1
& 0,\;0
& 0
\\
\mbox{Pure exotic}
& (0,\pm3/\sqrt{13})
& 1,\;\frac1{13}
& 0,\;-\frac4{13}
& 5
\\
\mbox{Mixed}
& (\pm\frac14,\pm\frac34)
& 1,\;-\frac18
& 0,\;-\frac38
& 6
\\
\hline
\end{array}
\)
\end{center}
%-------------------------------------------------------
Notice that for all the fixed points for which $\mathfrak{so}(6)$ is broken to $\mathfrak{so}(5)$ there are (at least) five transverse directions with zero anomalous dimensions corresponding to the tilt operator.
Additionally, the only IR stable point is the half-BPS line. The new fixed points (pure exotic and mixed) have both marginal and relevant deformations and are unstable.
 An overall sign change $(x,y)\rightarrow (-x,-y)$  corresponds to  changing the direction $n$ to $-n$, while the transformation $(x,y)\rightarrow (x,-y)$ is associated to a symmetry of the DCFT when the representation of the Wilson line is real.
%Besides the ordinary line, the fixed points are:
%\begin{itemize}
%\item the half-BPS lines $(x,y)=(\pm1,0)$, which are IR stable;
%\item the pure exotic lines $(0,\pm 3/\sqrt{13})$, which are IR stable within this plane;
%\item the mixed fixed points $(\pm\tfrac14,\pm\tfrac34)$, which have one relevant direction and separate the two basins along the lines $y=\pm3x$.
%\end{itemize}
%Outside this plane, the exotic lines are unstable: the $SO(5)$-vector deformation along $x^{M\neq6}$ has $\Delta-1=-\tfrac{4}{13}\hat{\lambda}_{\mathfrak{g}}$. A generic deformation of the ordinary line therefore flows to the half-BPS line.

\paragraph{Expansion around the half-BPS line.}
We now show how to recover the result for the beta-function around the half-BPS line at one loop starting from \eqref{betaordinaryfinal}. To do so, we eliminate the irrelevant longitudinal mode from the flow.
 Since \eqref{betaordinaryfinal} holds for $\xi=O(1)$, it can be expanded around $\xi^{(i,M)}=\delta^{i0}\delta^{M6}$, where it must reproduce the results of the main text. We write
\begin{equation}
\xi^{(i,6)}=\delta^{i0}+u^i\,,\qquad \xi^{(i,I)}=v^{(i,I)}\,,\qquad I=1,\dots,5\,,
\end{equation}
so that $u^i$ couples to $O_1^{(i)}(\Phi^6)$ and $v^{(i,I)}$ to the $D_1$-type operators of \eqref{deformation}. We use
$
\label{S00jk}
\mathsf{S}_{00ij}=-{\frac{h^\vee}{2}}\,\mathsf{g}_{ij}\,,
$
which follows from ${k^{bc}\rho_{\mathcal{R}}(T_b)\,O^{(i)}_{1a}\,\rho_{\mathcal{R}}(T_c)=\big(C_{\mathcal{R}}-\frac{h^\vee}{2}\big)O^{(i)}_{1a}}$, where $k^{bc}\rho_{\mathcal{R}}(T_b)\rho_{\mathcal{R}}(T_c)=C_{\mathcal{R}}\mathds{1}_{\mathcal{R}}$. 

At linear order, $\beta^{(i,6)}=\hat{\lambda}_{\mathfrak{g}}\,u^i$ and $\beta^{(i,I)}=0$: all the $O_1^{(i)}(\Phi^6)$ have $\Delta=1+\hat{\lambda}_{\mathfrak{g}}+\dots$, in agreement with \eqref{AndimAppendixResult}, while the $D_1$-type operators are marginal. 
At the next order the transverse couplings source the radial ones,
\begin{equation}
\beta^{(i,6)}=\hat{\lambda}_{\mathfrak{g}}\Big[u^i-{\tfrac{1}{h^\vee}}\,\mathsf{g}^{ii'}\mathsf{S}_{0i'jk}\,(v^j\cdot v^k)+{O(u^2,uv^2)}\Big]+\dots\,,
\end{equation}
where $\cdot$ is now the $SO(5)$ scalar product. {Notice that we neglect terms of order $O(u^2)$, which is justified because the longitudinal deformations are irrelevant already at linear order, whereas the transverse deformations are marginal. Consequently, the longitudinal couplings relax much faster than the transverse ones, and the flow approaches a slow manifold on which $u=O(v^2)$. Physically the slow manifold is exactly the space of half-BPS lines and their exotic deformations that was studied in the main text. To leading order, this manifold is obtained by setting the longitudinal beta-functions to zero, giving}
\begin{equation}
u^i={\tfrac{1}{h^\vee}}\,\mathsf{g}^{ii'}\mathsf{S}_{0i'jk}\,(v^j\cdot v^k)+\dots\,.
\end{equation}
For a pure super-displacement the definition of $\mathsf{S}_{00ij}$ gives $u^0=-\tfrac12|v^0|^2$, i.e.\ $\xi^{(0,6)}\simeq\sqrt{1-|v^0|^2}$, which is just a rotation of the half-BPS line. {Although the longitudinal directions have rapidly settled under the RG flow and lie on the slow manifold, there is still a residual flow in the transverse directions. This is described by the beta-function evaluated on the slow manifold, given by} inserting $u^i$ in the transverse components
\begin{equation}
\mathsf{g}_{ii'}\beta^{(i',I)}=-{\frac{\hat{\lambda}_{\mathfrak{g}}}{h^\vee}}\Big[\mathsf{S}_{jk\ell i}+{\tfrac{2}{h^\vee}}\,\mathsf{S}_{0m\ell i}\,\mathsf{g}^{mm'}\mathsf{S}_{0m'jk}\Big]v^{(\ell,I)}(v^j\cdot v^k)+O(v^5)\,,
\end{equation}
where the second term {arises from mixing with the irrelevant operators $O_1^{(m)}(\Phi^6)$ {in perturbation theory}. By using the methods described in Supplemental Materials II, it is possible to show the identity $\mathsf{S}_{0ijk}=\mu_{ijk}$, which in combination with \eqref{C2andL} and \eqref{cXandmumu} allows us to write the beta-function as}
\begin{equation}
\mathsf{g}_{ii'}\beta^{(i',I)}=\frac{{2}\hat{\lambda}_{\mathfrak{g}}}{h^\vee}\,\bigl(C^{\,2}_{B_1[2,0]^{+}}\bigr)^{(0)}_{jk,\ell i}\,v^{(\ell,I)}(v^j\cdot v^k)+O(v^5)\,.
\end{equation}
This coincides with \eqref{betafunctionwithC} once the couplings are rescaled in the normalisation of \eqref{deformation}, ${\zeta^{\mathcal{I}}=\big(\hat{\lambda}_{\mathfrak{g}}/h^\vee\big)^{1/2}v^{\mathcal{I}}}$. 
%This also provides an independent check of the sign of \eqref{betafunctionwithC}: the exotic deformations are marginally irrelevant.

\putbib
\end{bibunit}


%\bibliography{letter/letter}  

\begin{thebibliography}{53}%
\makeatletter
\providecommand \@ifxundefined [1]{%
 \@ifx{#1\undefined}
}%
\providecommand \@ifnum [1]{%
 \ifnum #1\expandafter \@firstoftwo
 \else \expandafter \@secondoftwo
 \fi
}%
\providecommand \@ifx [1]{%
 \ifx #1\expandafter \@firstoftwo
 \else \expandafter \@secondoftwo
 \fi
}%
\providecommand \natexlab [1]{#1}%
\providecommand \enquote  [1]{``#1''}%
\providecommand \bibnamefont  [1]{#1}%
\providecommand \bibfnamefont [1]{#1}%
\providecommand \citenamefont [1]{#1}%
\providecommand \href@noop [0]{\@secondoftwo}%
\providecommand \href [0]{\begingroup \@sanitize@url \@href}%
\providecommand \@href[1]{\@@startlink{#1}\@@href}%
\providecommand \@@href[1]{\endgroup#1\@@endlink}%
\providecommand \@sanitize@url [0]{\catcode `\\12\catcode `\$12\catcode `\&12\catcode `\#12\catcode `\^12\catcode `\_12\catcode `\%12\relax}%
\providecommand \@@startlink[1]{}%
\providecommand \@@endlink[0]{}%
\providecommand \url  [0]{\begingroup\@sanitize@url \@url }%
\providecommand \@url [1]{\endgroup\@href {#1}{\urlprefix }}%
\providecommand \urlprefix  [0]{URL }%
\providecommand \Eprint [0]{\href }%
\providecommand \doibase [0]{http://dx.doi.org/}%
\providecommand \selectlanguage [0]{\@gobble}%
\providecommand \bibinfo  [0]{\@secondoftwo}%
\providecommand \bibfield  [0]{\@secondoftwo}%
\providecommand \translation [1]{[#1]}%
\providecommand \BibitemOpen [0]{}%
\providecommand \bibitemStop [0]{}%
\providecommand \bibitemNoStop [0]{.\EOS\space}%
\providecommand \EOS [0]{\spacefactor3000\relax}%
\providecommand \BibitemShut  [1]{\csname bibitem#1\endcsname}%
\let\auto@bib@innerbib\@empty
%</preamble>
\bibitem [{\citenamefont {Wilson}(1974)}]{PhysRevD.10.2445}%
  \BibitemOpen
  \bibfield  {author} {\bibinfo {author} {\bibfnamefont {K.~G.}\ \bibnamefont {Wilson}},\ }\href {\doibase 10.1103/PhysRevD.10.2445} {\bibfield  {journal} {\bibinfo  {journal} {Phys. Rev. D}\ }\textbf {\bibinfo {volume} {10}},\ \bibinfo {pages} {2445} (\bibinfo {year} {1974})}\BibitemShut {NoStop}%
\bibitem [{\citenamefont {Makeenko}\ and\ \citenamefont {Migdal}(1979)}]{Makeenko:1979pb}%
  \BibitemOpen
  \bibfield  {author} {\bibinfo {author} {\bibfnamefont {Y.~M.}\ \bibnamefont {Makeenko}}\ and\ \bibinfo {author} {\bibfnamefont {A.~A.}\ \bibnamefont {Migdal}},\ }\href {\doibase 10.1016/0370-2693(79)90131-X} {\bibfield  {journal} {\bibinfo  {journal} {Phys. Lett. B}\ }\textbf {\bibinfo {volume} {88}},\ \bibinfo {pages} {135} (\bibinfo {year} {1979})},\ \bibinfo {note} {[Erratum: Phys.Lett.B 89, 437 (1980)]}\BibitemShut {NoStop}%
\bibitem [{\citenamefont {Polyakov}(1980)}]{Polyakov:1980ca}%
  \BibitemOpen
  \bibfield  {author} {\bibinfo {author} {\bibfnamefont {A.~M.}\ \bibnamefont {Polyakov}},\ }\href {\doibase 10.1016/0550-3213(80)90507-6} {\bibfield  {journal} {\bibinfo  {journal} {Nucl. Phys. B}\ }\textbf {\bibinfo {volume} {164}},\ \bibinfo {pages} {171} (\bibinfo {year} {1980})}\BibitemShut {NoStop}%
\bibitem [{\citenamefont {Maldacena}(1998)}]{Maldacena:1998im}%
  \BibitemOpen
  \bibfield  {author} {\bibinfo {author} {\bibfnamefont {J.~M.}\ \bibnamefont {Maldacena}},\ }\href {\doibase 10.1103/PhysRevLett.80.4859} {\bibfield  {journal} {\bibinfo  {journal} {Phys. Rev. Lett.}\ }\textbf {\bibinfo {volume} {80}},\ \bibinfo {pages} {4859} (\bibinfo {year} {1998})},\ \Eprint {http://arxiv.org/abs/hep-th/9803002} {arXiv:hep-th/9803002} \BibitemShut {NoStop}%
\bibitem [{\citenamefont {Drukker}\ and\ \citenamefont {Kawamoto}(2006)}]{Drukker:2006xg}%
  \BibitemOpen
  \bibfield  {author} {\bibinfo {author} {\bibfnamefont {N.}~\bibnamefont {Drukker}}\ and\ \bibinfo {author} {\bibfnamefont {S.}~\bibnamefont {Kawamoto}},\ }\href {\doibase 10.1088/1126-6708/2006/07/024} {\bibfield  {journal} {\bibinfo  {journal} {JHEP}\ }\textbf {\bibinfo {volume} {07}},\ \bibinfo {pages} {024} (\bibinfo {year} {2006})},\ \Eprint {http://arxiv.org/abs/hep-th/0604124} {arXiv:hep-th/0604124} \BibitemShut {NoStop}%
\bibitem [{Note1()}]{Note1}%
  \BibitemOpen
  \bibinfo {note} {We use a definite-parity basis, with parity-even operators real and parity-odd ones imaginary. Their two-point functions are then positive/negative definite and the OPE coefficients are real with $C_{O_1 O_2 O_3}=s_1s_2s_3 C_{O_2 O_1 O_3}$, see \cite {Homrich:2019cbt,Qiao:2017xif}.}\BibitemShut {Stop}%
\bibitem [{\citenamefont {Cavagli{\`a}}\ \emph {et~al.}(2025)\citenamefont {Cavagli{\`a}}, \citenamefont {Gromov},\ and\ \citenamefont {Preti}}]{Cavaglia:2023mmu}%
  \BibitemOpen
  \bibfield  {author} {\bibinfo {author} {\bibfnamefont {A.}~\bibnamefont {Cavagli{\`a}}}, \bibinfo {author} {\bibfnamefont {N.}~\bibnamefont {Gromov}}, \ and\ \bibinfo {author} {\bibfnamefont {M.}~\bibnamefont {Preti}},\ }\href {\doibase 10.1007/JHEP02(2025)026} {\bibfield  {journal} {\bibinfo  {journal} {JHEP}\ }\textbf {\bibinfo {volume} {02}},\ \bibinfo {pages} {026} (\bibinfo {year} {2025})},\ \Eprint {http://arxiv.org/abs/2312.11604} {arXiv:2312.11604 [hep-th]} \BibitemShut {NoStop}%
\bibitem [{\citenamefont {Gomis}(2025)}]{Gomis:2025gzb}%
  \BibitemOpen
  \bibfield  {author} {\bibinfo {author} {\bibfnamefont {J.}~\bibnamefont {Gomis}},\ }\href@noop {} {\enquote {\bibinfo {title} {{The AdS/$\mathsf{C}$-$\mathsf{P}$-${\mathsf T}$ Correspondence}},}\ } (\bibinfo {year} {2025}),\ \Eprint {http://arxiv.org/abs/2507.12467} {arXiv:2507.12467 [hep-th]} \BibitemShut {NoStop}%
\bibitem [{Note2()}]{Note2}%
  \BibitemOpen
  \bibinfo {note} {See \cite {Ferrero:2023znz} for a review of the relevant supermultiplets.}\BibitemShut {Stop}%
\bibitem [{Note3()}]{Note3}%
  \BibitemOpen
  \bibinfo {note} {It is commutative exactly when the weight-space decomposition of $\protect \mathcal {R}$ is multiplicity-free. For $\protect \mathfrak {su}(N)$, this leaves only the k-th symmetric and a-th antisymmetric representations.}\BibitemShut {Stop}%
\bibitem [{\citenamefont {Kirillov}(2000)}]{Kirillov_family}%
  \BibitemOpen
  \bibfield  {author} {\bibinfo {author} {\bibfnamefont {A.~A.}\ \bibnamefont {Kirillov}},\ }\href {\doibase 10.1090/S1079-6762-00-00075-5} {\bibfield  {journal} {\bibinfo  {journal} {Electron. Res. Announc. Amer. Math. Soc.}\ }\textbf {\bibinfo {volume} {6}},\ \bibinfo {pages} {7} (\bibinfo {year} {2000})}\BibitemShut {NoStop}%
\bibitem [{\citenamefont {Hausel}(2024)}]{Hausel_avatars}%
  \BibitemOpen
  \bibfield  {author} {\bibinfo {author} {\bibfnamefont {T.}~\bibnamefont {Hausel}},\ }\href {\doibase 10.1073/pnas.2319341121} {\bibfield  {journal} {\bibinfo  {journal} {Proceedings of the National Academy of Sciences}\ }\textbf {\bibinfo {volume} {121}},\ \bibinfo {pages} {e2319341121} (\bibinfo {year} {2024})}\BibitemShut {NoStop}%
\bibitem [{\citenamefont {Drukker}\ and\ \citenamefont {Plefka}(2009)}]{Drukker:2009sf}%
  \BibitemOpen
  \bibfield  {author} {\bibinfo {author} {\bibfnamefont {N.}~\bibnamefont {Drukker}}\ and\ \bibinfo {author} {\bibfnamefont {J.}~\bibnamefont {Plefka}},\ }\href {\doibase 10.1088/1126-6708/2009/04/052} {\bibfield  {journal} {\bibinfo  {journal} {JHEP}\ }\textbf {\bibinfo {volume} {04}},\ \bibinfo {pages} {052} (\bibinfo {year} {2009})},\ \Eprint {http://arxiv.org/abs/0901.3653} {arXiv:0901.3653 [hep-th]} \BibitemShut {NoStop}%
\bibitem [{\citenamefont {Liendo}\ and\ \citenamefont {Meneghelli}(2017)}]{Liendo:2016ymz}%
  \BibitemOpen
  \bibfield  {author} {\bibinfo {author} {\bibfnamefont {P.}~\bibnamefont {Liendo}}\ and\ \bibinfo {author} {\bibfnamefont {C.}~\bibnamefont {Meneghelli}},\ }\href {\doibase 10.1007/JHEP01(2017)122} {\bibfield  {journal} {\bibinfo  {journal} {JHEP}\ }\textbf {\bibinfo {volume} {01}},\ \bibinfo {pages} {122} (\bibinfo {year} {2017})},\ \Eprint {http://arxiv.org/abs/1608.05126} {arXiv:1608.05126 [hep-th]} \BibitemShut {NoStop}%
\bibitem [{\citenamefont {Giombi}\ and\ \citenamefont {Komatsu}(2018)}]{Giombi:2018qox}%
  \BibitemOpen
  \bibfield  {author} {\bibinfo {author} {\bibfnamefont {S.}~\bibnamefont {Giombi}}\ and\ \bibinfo {author} {\bibfnamefont {S.}~\bibnamefont {Komatsu}},\ }\href {\doibase 10.1007/JHEP05(2018)109} {\bibfield  {journal} {\bibinfo  {journal} {JHEP}\ }\textbf {\bibinfo {volume} {05}},\ \bibinfo {pages} {109} (\bibinfo {year} {2018})},\ \bibinfo {note} {[Erratum: JHEP 11, 123 (2018)]},\ \Eprint {http://arxiv.org/abs/1802.05201} {arXiv:1802.05201 [hep-th]} \BibitemShut {NoStop}%
\bibitem [{\citenamefont {Cachazo}\ \emph {et~al.}(2002)\citenamefont {Cachazo}, \citenamefont {Douglas}, \citenamefont {Seiberg},\ and\ \citenamefont {Witten}}]{Cachazo:2002ry}%
  \BibitemOpen
  \bibfield  {author} {\bibinfo {author} {\bibfnamefont {F.}~\bibnamefont {Cachazo}}, \bibinfo {author} {\bibfnamefont {M.~R.}\ \bibnamefont {Douglas}}, \bibinfo {author} {\bibfnamefont {N.}~\bibnamefont {Seiberg}}, \ and\ \bibinfo {author} {\bibfnamefont {E.}~\bibnamefont {Witten}},\ }\href {\doibase 10.1088/1126-6708/2002/12/071} {\bibfield  {journal} {\bibinfo  {journal} {JHEP}\ }\textbf {\bibinfo {volume} {12}},\ \bibinfo {pages} {071} (\bibinfo {year} {2002})},\ \Eprint {http://arxiv.org/abs/hep-th/0211170} {arXiv:hep-th/0211170} \BibitemShut {NoStop}%
\bibitem [{\citenamefont {Grant}\ \emph {et~al.}(2008)\citenamefont {Grant}, \citenamefont {Grassi}, \citenamefont {Kim},\ and\ \citenamefont {Minwalla}}]{Grant:2008sk}%
  \BibitemOpen
  \bibfield  {author} {\bibinfo {author} {\bibfnamefont {L.}~\bibnamefont {Grant}}, \bibinfo {author} {\bibfnamefont {P.~A.}\ \bibnamefont {Grassi}}, \bibinfo {author} {\bibfnamefont {S.}~\bibnamefont {Kim}}, \ and\ \bibinfo {author} {\bibfnamefont {S.}~\bibnamefont {Minwalla}},\ }\href {\doibase 10.1088/1126-6708/2008/05/049} {\bibfield  {journal} {\bibinfo  {journal} {JHEP}\ }\textbf {\bibinfo {volume} {05}},\ \bibinfo {pages} {049} (\bibinfo {year} {2008})},\ \Eprint {http://arxiv.org/abs/0803.4183} {arXiv:0803.4183 [hep-th]} \BibitemShut {NoStop}%
\bibitem [{\citenamefont {Chang}\ and\ \citenamefont {Yin}(2013)}]{Chang:2013fba}%
  \BibitemOpen
  \bibfield  {author} {\bibinfo {author} {\bibfnamefont {C.-M.}\ \bibnamefont {Chang}}\ and\ \bibinfo {author} {\bibfnamefont {X.}~\bibnamefont {Yin}},\ }\href {\doibase 10.1103/PhysRevD.88.106005} {\bibfield  {journal} {\bibinfo  {journal} {Phys. Rev. D}\ }\textbf {\bibinfo {volume} {88}},\ \bibinfo {pages} {106005} (\bibinfo {year} {2013})},\ \Eprint {http://arxiv.org/abs/1305.6314} {arXiv:1305.6314 [hep-th]} \BibitemShut {NoStop}%
\bibitem [{Note4()}]{Note4}%
  \BibitemOpen
  \bibinfo {note} {Because of unitarity bounds the undeformed theory does not have relevant operators.}\BibitemShut {Stop}%
\bibitem [{\citenamefont {Green}\ \emph {et~al.}(2010)\citenamefont {Green}, \citenamefont {Komargodski}, \citenamefont {Seiberg}, \citenamefont {Tachikawa},\ and\ \citenamefont {Wecht}}]{Green:2010da}%
  \BibitemOpen
  \bibfield  {author} {\bibinfo {author} {\bibfnamefont {D.}~\bibnamefont {Green}}, \bibinfo {author} {\bibfnamefont {Z.}~\bibnamefont {Komargodski}}, \bibinfo {author} {\bibfnamefont {N.}~\bibnamefont {Seiberg}}, \bibinfo {author} {\bibfnamefont {Y.}~\bibnamefont {Tachikawa}}, \ and\ \bibinfo {author} {\bibfnamefont {B.}~\bibnamefont {Wecht}},\ }\href {\doibase 10.1007/JHEP06(2010)106} {\bibfield  {journal} {\bibinfo  {journal} {JHEP}\ }\textbf {\bibinfo {volume} {06}},\ \bibinfo {pages} {106} (\bibinfo {year} {2010})},\ \Eprint {http://arxiv.org/abs/1005.3546} {arXiv:1005.3546 [hep-th]} \BibitemShut {NoStop}%
\bibitem [{\citenamefont {Komargodski}\ and\ \citenamefont {Simmons-Duffin}(2017)}]{Komargodski:2016auf}%
  \BibitemOpen
  \bibfield  {author} {\bibinfo {author} {\bibfnamefont {Z.}~\bibnamefont {Komargodski}}\ and\ \bibinfo {author} {\bibfnamefont {D.}~\bibnamefont {Simmons-Duffin}},\ }\href {\doibase 10.1088/1751-8121/aa6087} {\bibfield  {journal} {\bibinfo  {journal} {J. Phys. A}\ }\textbf {\bibinfo {volume} {50}},\ \bibinfo {pages} {154001} (\bibinfo {year} {2017})},\ \Eprint {http://arxiv.org/abs/1603.04444} {arXiv:1603.04444 [hep-th]} \BibitemShut {NoStop}%
\bibitem [{\citenamefont {Behan}\ \emph {et~al.}(2017)\citenamefont {Behan}, \citenamefont {Rastelli}, \citenamefont {Rychkov},\ and\ \citenamefont {Zan}}]{Behan:2017emf}%
  \BibitemOpen
  \bibfield  {author} {\bibinfo {author} {\bibfnamefont {C.}~\bibnamefont {Behan}}, \bibinfo {author} {\bibfnamefont {L.}~\bibnamefont {Rastelli}}, \bibinfo {author} {\bibfnamefont {S.}~\bibnamefont {Rychkov}}, \ and\ \bibinfo {author} {\bibfnamefont {B.}~\bibnamefont {Zan}},\ }\href {\doibase 10.1088/1751-8121/aa8099} {\bibfield  {journal} {\bibinfo  {journal} {J. Phys. A}\ }\textbf {\bibinfo {volume} {50}},\ \bibinfo {pages} {354002} (\bibinfo {year} {2017})},\ \Eprint {http://arxiv.org/abs/1703.05325} {arXiv:1703.05325 [hep-th]} \BibitemShut {NoStop}%
\bibitem [{\citenamefont {Behan}(2018)}]{Behan:2017mwi}%
  \BibitemOpen
  \bibfield  {author} {\bibinfo {author} {\bibfnamefont {C.}~\bibnamefont {Behan}},\ }\href {\doibase 10.1007/JHEP03(2018)127} {\bibfield  {journal} {\bibinfo  {journal} {JHEP}\ }\textbf {\bibinfo {volume} {03}},\ \bibinfo {pages} {127} (\bibinfo {year} {2018})},\ \Eprint {http://arxiv.org/abs/1709.03967} {arXiv:1709.03967 [hep-th]} \BibitemShut {NoStop}%
\bibitem [{\citenamefont {Liendo}\ \emph {et~al.}(2018)\citenamefont {Liendo}, \citenamefont {Meneghelli},\ and\ \citenamefont {Mitev}}]{Liendo:2018ukf}%
  \BibitemOpen
  \bibfield  {author} {\bibinfo {author} {\bibfnamefont {P.}~\bibnamefont {Liendo}}, \bibinfo {author} {\bibfnamefont {C.}~\bibnamefont {Meneghelli}}, \ and\ \bibinfo {author} {\bibfnamefont {V.}~\bibnamefont {Mitev}},\ }\href {\doibase 10.1007/JHEP10(2018)077} {\bibfield  {journal} {\bibinfo  {journal} {JHEP}\ }\textbf {\bibinfo {volume} {10}},\ \bibinfo {pages} {077} (\bibinfo {year} {2018})},\ \Eprint {http://arxiv.org/abs/1806.01862} {arXiv:1806.01862 [hep-th]} \BibitemShut {NoStop}%
\bibitem [{Note5()}]{Note5}%
  \BibitemOpen
  \bibinfo {note} {Notice that $(C^{\protect \,2}_{B_1[2,0]})_{(ij)k\ell }=2(C^{\protect \,2}_{B_1[2,0]^+})_{ijk\ell }$ where $A_{(ij)}=A_{ij}+A_{ji}$.}\BibitemShut {Stop}%
\bibitem [{\citenamefont {Gabai}\ \emph {et~al.}(2025)\citenamefont {Gabai}, \citenamefont {Sever},\ and\ \citenamefont {Zhong}}]{Gabai:2025zcs}%
  \BibitemOpen
  \bibfield  {author} {\bibinfo {author} {\bibfnamefont {B.}~\bibnamefont {Gabai}}, \bibinfo {author} {\bibfnamefont {A.}~\bibnamefont {Sever}}, \ and\ \bibinfo {author} {\bibfnamefont {D.-l.}\ \bibnamefont {Zhong}},\ }\href {\doibase 10.1103/gsfg-wrps} {\bibfield  {journal} {\bibinfo  {journal} {Phys. Rev. D}\ }\textbf {\bibinfo {volume} {112}},\ \bibinfo {pages} {065004} (\bibinfo {year} {2025})},\ \Eprint {http://arxiv.org/abs/2501.06900} {arXiv:2501.06900 [hep-th]} \BibitemShut {NoStop}%
\bibitem [{\citenamefont {Girault}\ \emph {et~al.}(2025)\citenamefont {Girault}, \citenamefont {Paulos},\ and\ \citenamefont {van Vliet}}]{Girault:2025kzt}%
  \BibitemOpen
  \bibfield  {author} {\bibinfo {author} {\bibfnamefont {B.}~\bibnamefont {Girault}}, \bibinfo {author} {\bibfnamefont {M.~F.}\ \bibnamefont {Paulos}}, \ and\ \bibinfo {author} {\bibfnamefont {P.}~\bibnamefont {van Vliet}},\ }\href@noop {} {\  (\bibinfo {year} {2025})},\ \Eprint {http://arxiv.org/abs/2509.26561} {arXiv:2509.26561 [hep-th]} \BibitemShut {NoStop}%
\bibitem [{\citenamefont {Drukker}\ \emph {et~al.}(2025)\citenamefont {Drukker}, \citenamefont {Kong},\ and\ \citenamefont {Kravchuk}}]{Drukker:2025dfm}%
  \BibitemOpen
  \bibfield  {author} {\bibinfo {author} {\bibfnamefont {N.}~\bibnamefont {Drukker}}, \bibinfo {author} {\bibfnamefont {Z.}~\bibnamefont {Kong}}, \ and\ \bibinfo {author} {\bibfnamefont {P.}~\bibnamefont {Kravchuk}},\ }\href@noop {} {\  (\bibinfo {year} {2025})},\ \Eprint {http://arxiv.org/abs/2512.15913} {arXiv:2512.15913 [hep-th]} \BibitemShut {NoStop}%
\bibitem [{\citenamefont {Belton}\ and\ \citenamefont {Kong}(2026)}]{Belton:2025ief}%
  \BibitemOpen
  \bibfield  {author} {\bibinfo {author} {\bibfnamefont {J.}~\bibnamefont {Belton}}\ and\ \bibinfo {author} {\bibfnamefont {Z.}~\bibnamefont {Kong}},\ }\href {\doibase 10.1007/JHEP05(2026)103} {\bibfield  {journal} {\bibinfo  {journal} {JHEP}\ }\textbf {\bibinfo {volume} {05}},\ \bibinfo {pages} {103} (\bibinfo {year} {2026})},\ \Eprint {http://arxiv.org/abs/2510.08519} {arXiv:2510.08519 [hep-th]} \BibitemShut {NoStop}%
\bibitem [{Note6()}]{Note6}%
  \BibitemOpen
  \bibinfo {note} {Equivalently this can be rewritten as $C_{0i\protect \mathcal {O}}+C_{i0\protect \mathcal {O}}=0$ $\forall $ $\protect \mathcal {O}$ of type $B_1[2,0]$.}\BibitemShut {Stop}%
\bibitem [{Note7()}]{Note7}%
  \BibitemOpen
  \bibinfo {note} {A nontrivial defect conformal manifold is not excluded, but would require relations in the quarter-BPS chiral ring.}\BibitemShut {Stop}%
\bibitem [{\citenamefont {Ferrero}\ and\ \citenamefont {Meneghelli}(2021)}]{Ferrero:2021bsb}%
  \BibitemOpen
  \bibfield  {author} {\bibinfo {author} {\bibfnamefont {P.}~\bibnamefont {Ferrero}}\ and\ \bibinfo {author} {\bibfnamefont {C.}~\bibnamefont {Meneghelli}},\ }\href {\doibase 10.1103/PhysRevD.104.L081703} {\bibfield  {journal} {\bibinfo  {journal} {Phys. Rev. D}\ }\textbf {\bibinfo {volume} {104}},\ \bibinfo {pages} {L081703} (\bibinfo {year} {2021})},\ \Eprint {http://arxiv.org/abs/2103.10440} {arXiv:2103.10440 [hep-th]} \BibitemShut {NoStop}%
\bibitem [{Note8()}]{Note8}%
  \BibitemOpen
  \bibinfo {note} {The integral thus reduces to boundary terms determined by OPE coefficients. See \cite {Chen:2026vml} for a similar mechanism.}\BibitemShut {Stop}%
\bibitem [{Note9()}]{Note9}%
  \BibitemOpen
  \bibinfo {note} {The limits $g_{YM}^2\to 0$ and $a\to 0$ do not commute because of terms $a^{\Delta (g_{YM}^2)-1}$ with $\Delta (0)=1$. This is resolved by including perturbative mixing between $\Phi ^I$ and $\Phi ^6$; see also~\cite {Cavaglia:2022qpg,Gabai:2025zcs}.}\BibitemShut {Stop}%
\bibitem [{\citenamefont {Artico}\ \emph {et~al.}()\citenamefont {Artico}, \citenamefont {Meneghelli}, \citenamefont {Savi},\ and\ \citenamefont {Treilis}}]{Artico:2026tba}%
  \BibitemOpen
  \bibfield  {author} {\bibinfo {author} {\bibfnamefont {D.}~\bibnamefont {Artico}}, \bibinfo {author} {\bibfnamefont {C.}~\bibnamefont {Meneghelli}}, \bibinfo {author} {\bibfnamefont {M.}~\bibnamefont {Savi}}, \ and\ \bibinfo {author} {\bibfnamefont {R.}~\bibnamefont {Treilis}},\ }\href@noop {} {\ }\Eprint {http://arxiv.org/abs/to appear} {to appear} \BibitemShut {NoStop}%
\bibitem [{\citenamefont {Artico}\ \emph {et~al.}(2025)\citenamefont {Artico}, \citenamefont {Barrat},\ and\ \citenamefont {Peveri}}]{Artico:2024wut}%
  \BibitemOpen
  \bibfield  {author} {\bibinfo {author} {\bibfnamefont {D.}~\bibnamefont {Artico}}, \bibinfo {author} {\bibfnamefont {J.}~\bibnamefont {Barrat}}, \ and\ \bibinfo {author} {\bibfnamefont {G.}~\bibnamefont {Peveri}},\ }\href {\doibase 10.1007/JHEP02(2025)190} {\bibfield  {journal} {\bibinfo  {journal} {JHEP}\ }\textbf {\bibinfo {volume} {02}},\ \bibinfo {pages} {190} (\bibinfo {year} {2025})},\ \Eprint {http://arxiv.org/abs/2410.08271} {arXiv:2410.08271 [hep-th]} \BibitemShut {NoStop}%
\bibitem [{Note10()}]{Note10}%
  \BibitemOpen
  \bibinfo {note} {A similar multiplicity of displacement-type multiplets occurs for half-BPS defects in the $6d$ $(2,0)$ theory; see \cite {Meneghelli:2022gps}.}\BibitemShut {Stop}%
\bibitem [{\citenamefont {Polchinski}\ and\ \citenamefont {Sully}(2011)}]{Polchinski:2011im}%
  \BibitemOpen
  \bibfield  {author} {\bibinfo {author} {\bibfnamefont {J.}~\bibnamefont {Polchinski}}\ and\ \bibinfo {author} {\bibfnamefont {J.}~\bibnamefont {Sully}},\ }\href {\doibase 10.1007/JHEP10(2011)059} {\bibfield  {journal} {\bibinfo  {journal} {JHEP}\ }\textbf {\bibinfo {volume} {10}},\ \bibinfo {pages} {059} (\bibinfo {year} {2011})},\ \Eprint {http://arxiv.org/abs/1104.5077} {arXiv:1104.5077 [hep-th]} \BibitemShut {NoStop}%
\bibitem [{\citenamefont {Beccaria}\ \emph {et~al.}(2022)\citenamefont {Beccaria}, \citenamefont {Giombi},\ and\ \citenamefont {Tseytlin}}]{Beccaria:2022bcr}%
  \BibitemOpen
  \bibfield  {author} {\bibinfo {author} {\bibfnamefont {M.}~\bibnamefont {Beccaria}}, \bibinfo {author} {\bibfnamefont {S.}~\bibnamefont {Giombi}}, \ and\ \bibinfo {author} {\bibfnamefont {A.~A.}\ \bibnamefont {Tseytlin}},\ }\href {\doibase 10.1088/1751-8121/ac7018} {\bibfield  {journal} {\bibinfo  {journal} {J. Phys. A}\ }\textbf {\bibinfo {volume} {55}},\ \bibinfo {pages} {255401} (\bibinfo {year} {2022})},\ \Eprint {http://arxiv.org/abs/2202.00028} {arXiv:2202.00028 [hep-th]} \BibitemShut {NoStop}%
\bibitem [{Note11()}]{Note11}%
  \BibitemOpen
  \bibinfo {note} {These operators may become more relevant at stronger coupling. For the fundamental ordinary line in the planar limit, their scaling dimension tends to zero at large 't Hooft coupling~\cite {Beccaria:2017rbe}.}\BibitemShut {Stop}%
\bibitem [{\citenamefont {Aharony}\ \emph {et~al.}(2023)\citenamefont {Aharony}, \citenamefont {Cuomo}, \citenamefont {Komargodski}, \citenamefont {Mezei},\ and\ \citenamefont {Raviv-Moshe}}]{Aharony:2022ntz}%
  \BibitemOpen
  \bibfield  {author} {\bibinfo {author} {\bibfnamefont {O.}~\bibnamefont {Aharony}}, \bibinfo {author} {\bibfnamefont {G.}~\bibnamefont {Cuomo}}, \bibinfo {author} {\bibfnamefont {Z.}~\bibnamefont {Komargodski}}, \bibinfo {author} {\bibfnamefont {M.}~\bibnamefont {Mezei}}, \ and\ \bibinfo {author} {\bibfnamefont {A.}~\bibnamefont {Raviv-Moshe}},\ }\href {\doibase 10.1103/PhysRevLett.130.151601} {\bibfield  {journal} {\bibinfo  {journal} {Phys. Rev. Lett.}\ }\textbf {\bibinfo {volume} {130}},\ \bibinfo {pages} {151601} (\bibinfo {year} {2023})},\ \Eprint {http://arxiv.org/abs/2211.11775} {arXiv:2211.11775 [hep-th]} \BibitemShut {NoStop}%
\bibitem [{\citenamefont {Chang}\ and\ \citenamefont {Lin}(2024)}]{Chang:2024zqi}%
  \BibitemOpen
  \bibfield  {author} {\bibinfo {author} {\bibfnamefont {C.-M.}\ \bibnamefont {Chang}}\ and\ \bibinfo {author} {\bibfnamefont {Y.-H.}\ \bibnamefont {Lin}},\ }\href@noop {} {\  (\bibinfo {year} {2024})},\ \Eprint {http://arxiv.org/abs/2402.10129} {arXiv:2402.10129 [hep-th]} \BibitemShut {NoStop}%
\bibitem [{\citenamefont {Bonetti}\ and\ \citenamefont {Meneghelli}(2025)}]{Bonetti:2025kan}%
  \BibitemOpen
  \bibfield  {author} {\bibinfo {author} {\bibfnamefont {F.}~\bibnamefont {Bonetti}}\ and\ \bibinfo {author} {\bibfnamefont {C.}~\bibnamefont {Meneghelli}},\ }\href@noop {} {\  (\bibinfo {year} {2025})},\ \Eprint {http://arxiv.org/abs/2506.15678} {arXiv:2506.15678 [hep-th]} \BibitemShut {NoStop}%
\bibitem [{Note12()}]{Note12}%
  \BibitemOpen
  \bibinfo {note} {For a recent study of the relation between continuous families of CFTs and conformal manifolds see \cite {Komatsu:2025cai}.}\BibitemShut {Stop}%
\bibitem [{\citenamefont {Gomis}\ and\ \citenamefont {Passerini}(2006)}]{Gomis:2006sb}%
  \BibitemOpen
  \bibfield  {author} {\bibinfo {author} {\bibfnamefont {J.}~\bibnamefont {Gomis}}\ and\ \bibinfo {author} {\bibfnamefont {F.}~\bibnamefont {Passerini}},\ }\href {\doibase 10.1088/1126-6708/2006/08/074} {\bibfield  {journal} {\bibinfo  {journal} {JHEP}\ }\textbf {\bibinfo {volume} {08}},\ \bibinfo {pages} {074} (\bibinfo {year} {2006})},\ \Eprint {http://arxiv.org/abs/hep-th/0604007} {arXiv:hep-th/0604007} \BibitemShut {NoStop}%
\bibitem [{\citenamefont {Homrich}\ \emph {et~al.}(2019)\citenamefont {Homrich}, \citenamefont {Penedones}, \citenamefont {Toledo}, \citenamefont {van Rees},\ and\ \citenamefont {Vieira}}]{Homrich:2019cbt}%
  \BibitemOpen
  \bibfield  {author} {\bibinfo {author} {\bibfnamefont {A.}~\bibnamefont {Homrich}}, \bibinfo {author} {\bibfnamefont {J.}~\bibnamefont {Penedones}}, \bibinfo {author} {\bibfnamefont {J.}~\bibnamefont {Toledo}}, \bibinfo {author} {\bibfnamefont {B.~C.}\ \bibnamefont {van Rees}}, \ and\ \bibinfo {author} {\bibfnamefont {P.}~\bibnamefont {Vieira}},\ }\href {\doibase 10.1007/JHEP11(2019)076} {\bibfield  {journal} {\bibinfo  {journal} {JHEP}\ }\textbf {\bibinfo {volume} {11}},\ \bibinfo {pages} {076} (\bibinfo {year} {2019})},\ \Eprint {http://arxiv.org/abs/1905.06905} {arXiv:1905.06905 [hep-th]} \BibitemShut {NoStop}%
\bibitem [{\citenamefont {Qiao}\ and\ \citenamefont {Rychkov}(2017)}]{Qiao:2017xif}%
  \BibitemOpen
  \bibfield  {author} {\bibinfo {author} {\bibfnamefont {J.}~\bibnamefont {Qiao}}\ and\ \bibinfo {author} {\bibfnamefont {S.}~\bibnamefont {Rychkov}},\ }\href {\doibase 10.1007/JHEP12(2017)119} {\bibfield  {journal} {\bibinfo  {journal} {JHEP}\ }\textbf {\bibinfo {volume} {12}},\ \bibinfo {pages} {119} (\bibinfo {year} {2017})},\ \Eprint {http://arxiv.org/abs/1709.00008} {arXiv:1709.00008 [hep-th]} \BibitemShut {NoStop}%
\bibitem [{\citenamefont {Ferrero}\ and\ \citenamefont {Meneghelli}(2024)}]{Ferrero:2023znz}%
  \BibitemOpen
  \bibfield  {author} {\bibinfo {author} {\bibfnamefont {P.}~\bibnamefont {Ferrero}}\ and\ \bibinfo {author} {\bibfnamefont {C.}~\bibnamefont {Meneghelli}},\ }\href {\doibase 10.1007/JHEP05(2024)090} {\bibfield  {journal} {\bibinfo  {journal} {JHEP}\ }\textbf {\bibinfo {volume} {05}},\ \bibinfo {pages} {090} (\bibinfo {year} {2024})},\ \Eprint {http://arxiv.org/abs/2312.12550} {arXiv:2312.12550 [hep-th]} \BibitemShut {NoStop}%
\bibitem [{\citenamefont {Chen}\ \emph {et~al.}(2026)\citenamefont {Chen}, \citenamefont {Colin-Ellerin}, \citenamefont {Mamroud},\ and\ \citenamefont {Papadodimas}}]{Chen:2026vml}%
  \BibitemOpen
  \bibfield  {author} {\bibinfo {author} {\bibfnamefont {Y.}~\bibnamefont {Chen}}, \bibinfo {author} {\bibfnamefont {S.}~\bibnamefont {Colin-Ellerin}}, \bibinfo {author} {\bibfnamefont {O.}~\bibnamefont {Mamroud}}, \ and\ \bibinfo {author} {\bibfnamefont {K.}~\bibnamefont {Papadodimas}},\ }\href@noop {} {\  (\bibinfo {year} {2026})},\ \Eprint {http://arxiv.org/abs/2604.23287} {arXiv:2604.23287 [hep-th]} \BibitemShut {NoStop}%
\bibitem [{\citenamefont {Cavagli{\`a}}\ \emph {et~al.}(2022)\citenamefont {Cavagli{\`a}}, \citenamefont {Gromov}, \citenamefont {Julius},\ and\ \citenamefont {Preti}}]{Cavaglia:2022qpg}%
  \BibitemOpen
  \bibfield  {author} {\bibinfo {author} {\bibfnamefont {A.}~\bibnamefont {Cavagli{\`a}}}, \bibinfo {author} {\bibfnamefont {N.}~\bibnamefont {Gromov}}, \bibinfo {author} {\bibfnamefont {J.}~\bibnamefont {Julius}}, \ and\ \bibinfo {author} {\bibfnamefont {M.}~\bibnamefont {Preti}},\ }\href {\doibase 10.1007/JHEP05(2022)164} {\bibfield  {journal} {\bibinfo  {journal} {JHEP}\ }\textbf {\bibinfo {volume} {05}},\ \bibinfo {pages} {164} (\bibinfo {year} {2022})},\ \Eprint {http://arxiv.org/abs/2203.09556} {arXiv:2203.09556 [hep-th]} \BibitemShut {NoStop}%
\bibitem [{\citenamefont {Meneghelli}\ and\ \citenamefont {Tr{\'e}panier}(2023)}]{Meneghelli:2022gps}%
  \BibitemOpen
  \bibfield  {author} {\bibinfo {author} {\bibfnamefont {C.}~\bibnamefont {Meneghelli}}\ and\ \bibinfo {author} {\bibfnamefont {M.}~\bibnamefont {Tr{\'e}panier}},\ }\href {\doibase 10.1007/JHEP07(2023)165} {\bibfield  {journal} {\bibinfo  {journal} {JHEP}\ }\textbf {\bibinfo {volume} {07}},\ \bibinfo {pages} {165} (\bibinfo {year} {2023})},\ \Eprint {http://arxiv.org/abs/2212.05020} {arXiv:2212.05020 [hep-th]} \BibitemShut {NoStop}%
\bibitem [{\citenamefont {Beccaria}\ \emph {et~al.}(2018)\citenamefont {Beccaria}, \citenamefont {Giombi},\ and\ \citenamefont {Tseytlin}}]{Beccaria:2017rbe}%
  \BibitemOpen
  \bibfield  {author} {\bibinfo {author} {\bibfnamefont {M.}~\bibnamefont {Beccaria}}, \bibinfo {author} {\bibfnamefont {S.}~\bibnamefont {Giombi}}, \ and\ \bibinfo {author} {\bibfnamefont {A.}~\bibnamefont {Tseytlin}},\ }\href {\doibase 10.1007/JHEP03(2018)131} {\bibfield  {journal} {\bibinfo  {journal} {JHEP}\ }\textbf {\bibinfo {volume} {03}},\ \bibinfo {pages} {131} (\bibinfo {year} {2018})},\ \Eprint {http://arxiv.org/abs/1712.06874} {arXiv:1712.06874 [hep-th]} \BibitemShut {NoStop}%
\bibitem [{\citenamefont {Komatsu}\ \emph {et~al.}(2025)\citenamefont {Komatsu}, \citenamefont {Kusuki}, \citenamefont {Meineri},\ and\ \citenamefont {Ooguri}}]{Komatsu:2025cai}%
  \BibitemOpen
  \bibfield  {author} {\bibinfo {author} {\bibfnamefont {S.}~\bibnamefont {Komatsu}}, \bibinfo {author} {\bibfnamefont {Y.}~\bibnamefont {Kusuki}}, \bibinfo {author} {\bibfnamefont {M.}~\bibnamefont {Meineri}}, \ and\ \bibinfo {author} {\bibfnamefont {H.}~\bibnamefont {Ooguri}},\ }\href@noop {} {\  (\bibinfo {year} {2025})},\ \Eprint {http://arxiv.org/abs/2512.11045} {arXiv:2512.11045 [hep-th]} \BibitemShut {NoStop}%
\end{thebibliography}%


\begin{thebibliography}{19}%
\makeatletter
\providecommand \@ifxundefined [1]{%
 \@ifx{#1\undefined}
}%
\providecommand \@ifnum [1]{%
 \ifnum #1\expandafter \@firstoftwo
 \else \expandafter \@secondoftwo
 \fi
}%
\providecommand \@ifx [1]{%
 \ifx #1\expandafter \@firstoftwo
 \else \expandafter \@secondoftwo
 \fi
}%
\providecommand \natexlab [1]{#1}%
\providecommand \enquote  [1]{``#1''}%
\providecommand \bibnamefont  [1]{#1}%
\providecommand \bibfnamefont [1]{#1}%
\providecommand \citenamefont [1]{#1}%
\providecommand \href@noop [0]{\@secondoftwo}%
\providecommand \href [0]{\begingroup \@sanitize@url \@href}%
\providecommand \@href[1]{\@@startlink{#1}\@@href}%
\providecommand \@@href[1]{\endgroup#1\@@endlink}%
\providecommand \@sanitize@url [0]{\catcode `\\12\catcode `\$12\catcode `\&12\catcode `\#12\catcode `\^12\catcode `\_12\catcode `\%12\relax}%
\providecommand \@@startlink[1]{}%
\providecommand \@@endlink[0]{}%
\providecommand \url  [0]{\begingroup\@sanitize@url \@url }%
\providecommand \@url [1]{\endgroup\@href {#1}{\urlprefix }}%
\providecommand \urlprefix  [0]{URL }%
\providecommand \Eprint [0]{\href }%
\providecommand \doibase [0]{http://dx.doi.org/}%
\providecommand \selectlanguage [0]{\@gobble}%
\providecommand \bibinfo  [0]{\@secondoftwo}%
\providecommand \bibfield  [0]{\@secondoftwo}%
\providecommand \translation [1]{[#1]}%
\providecommand \BibitemOpen [0]{}%
\providecommand \bibitemStop [0]{}%
\providecommand \bibitemNoStop [0]{.\EOS\space}%
\providecommand \EOS [0]{\spacefactor3000\relax}%
\providecommand \BibitemShut  [1]{\csname bibitem#1\endcsname}%
\let\auto@bib@innerbib\@empty
%</preamble>
\bibitem [{\citenamefont {Okubo}\ and\ \citenamefont {Myung}(1981)}]{Okubo:1981}%
  \BibitemOpen
  \bibfield  {author} {\bibinfo {author} {\bibfnamefont {S.}~\bibnamefont {Okubo}}\ and\ \bibinfo {author} {\bibfnamefont {H.}~\bibnamefont {Myung}},\ }\href@noop {} {\bibfield  {journal} {\bibinfo  {journal} {Hadronic J.}\ }\textbf {\bibinfo {volume} {4}},\ \bibinfo {pages} {199} (\bibinfo {year} {1981})}\BibitemShut {NoStop}%
\bibitem [{\citenamefont {Perelomov}\ and\ \citenamefont {Popov}(1968)}]{PP1968}%
  \BibitemOpen
  \bibfield  {author} {\bibinfo {author} {\bibfnamefont {A.~M.}\ \bibnamefont {Perelomov}}\ and\ \bibinfo {author} {\bibfnamefont {V.~S.}\ \bibnamefont {Popov}},\ }\href@noop {} {\bibfield  {journal} {\bibinfo  {journal} {Izv. Akad. Nauk SSSR Ser. Mat.}\ }\textbf {\bibinfo {volume} {32}},\ \bibinfo {pages} {1368} (\bibinfo {year} {1968})}\BibitemShut {NoStop}%
\bibitem [{\citenamefont {Molev}\ \emph {et~al.}(1994)\citenamefont {Molev}, \citenamefont {Nazarov},\ and\ \citenamefont {Olshanskii}}]{Molev:1994}%
  \BibitemOpen
  \bibfield  {author} {\bibinfo {author} {\bibfnamefont {A.}~\bibnamefont {Molev}}, \bibinfo {author} {\bibfnamefont {M.}~\bibnamefont {Nazarov}}, \ and\ \bibinfo {author} {\bibfnamefont {G.}~\bibnamefont {Olshanskii}},\ }\href {https://arxiv.org/abs/hep-th/9409025} {\enquote {\bibinfo {title} {Yangians and classical lie algebras},}\ } (\bibinfo {year} {1994}),\ \Eprint {http://arxiv.org/abs/hep-th/9409025} {arXiv:hep-th/9409025 [hep-th]} \BibitemShut {NoStop}%
\bibitem [{\citenamefont {Beisert}\ \emph {et~al.}(2003)\citenamefont {Beisert}, \citenamefont {Kristjansen}, \citenamefont {Plefka}, \citenamefont {Semenoff},\ and\ \citenamefont {Staudacher}}]{Beisert:2002bb}%
  \BibitemOpen
  \bibfield  {author} {\bibinfo {author} {\bibfnamefont {N.}~\bibnamefont {Beisert}}, \bibinfo {author} {\bibfnamefont {C.}~\bibnamefont {Kristjansen}}, \bibinfo {author} {\bibfnamefont {J.}~\bibnamefont {Plefka}}, \bibinfo {author} {\bibfnamefont {G.~W.}\ \bibnamefont {Semenoff}}, \ and\ \bibinfo {author} {\bibfnamefont {M.}~\bibnamefont {Staudacher}},\ }\href {\doibase 10.1016/S0550-3213(02)01025-8} {\bibfield  {journal} {\bibinfo  {journal} {Nucl. Phys. B}\ }\textbf {\bibinfo {volume} {650}},\ \bibinfo {pages} {125} (\bibinfo {year} {2003})},\ \Eprint {http://arxiv.org/abs/hep-th/0208178} {arXiv:hep-th/0208178} \BibitemShut {NoStop}%
\bibitem [{\citenamefont {Drukker}\ and\ \citenamefont {Plefka}(2009)}]{Drukker:2008pi}%
  \BibitemOpen
  \bibfield  {author} {\bibinfo {author} {\bibfnamefont {N.}~\bibnamefont {Drukker}}\ and\ \bibinfo {author} {\bibfnamefont {J.}~\bibnamefont {Plefka}},\ }\href {\doibase 10.1088/1126-6708/2009/04/001} {\bibfield  {journal} {\bibinfo  {journal} {JHEP}\ }\textbf {\bibinfo {volume} {04}},\ \bibinfo {pages} {001} (\bibinfo {year} {2009})},\ \Eprint {http://arxiv.org/abs/0812.3341} {arXiv:0812.3341 [hep-th]} \BibitemShut {NoStop}%
\bibitem [{\citenamefont {Usyukina}\ and\ \citenamefont {Davydychev}(1994)}]{Usyukina:1994iw}%
  \BibitemOpen
  \bibfield  {author} {\bibinfo {author} {\bibfnamefont {N.~I.}\ \bibnamefont {Usyukina}}\ and\ \bibinfo {author} {\bibfnamefont {A.~I.}\ \bibnamefont {Davydychev}},\ }\href {\doibase 10.1016/0370-2693(94)90874-5} {\bibfield  {journal} {\bibinfo  {journal} {Phys. Lett. B}\ }\textbf {\bibinfo {volume} {332}},\ \bibinfo {pages} {159} (\bibinfo {year} {1994})},\ \Eprint {http://arxiv.org/abs/hep-ph/9402223} {arXiv:hep-ph/9402223} \BibitemShut {NoStop}%
\bibitem [{\citenamefont {Usyukina}\ and\ \citenamefont {Davydychev}(1995)}]{Usyukina:1994eg}%
  \BibitemOpen
  \bibfield  {author} {\bibinfo {author} {\bibfnamefont {N.~I.}\ \bibnamefont {Usyukina}}\ and\ \bibinfo {author} {\bibfnamefont {A.~I.}\ \bibnamefont {Davydychev}},\ }\href {\doibase 10.1016/0370-2693(95)00136-9} {\bibfield  {journal} {\bibinfo  {journal} {Phys. Lett. B}\ }\textbf {\bibinfo {volume} {348}},\ \bibinfo {pages} {503} (\bibinfo {year} {1995})},\ \Eprint {http://arxiv.org/abs/hep-ph/9412356} {arXiv:hep-ph/9412356} \BibitemShut {NoStop}%
\bibitem [{\citenamefont {Liendo}\ \emph {et~al.}(2018)\citenamefont {Liendo}, \citenamefont {Meneghelli},\ and\ \citenamefont {Mitev}}]{Liendo:2018ukf}%
  \BibitemOpen
  \bibfield  {author} {\bibinfo {author} {\bibfnamefont {P.}~\bibnamefont {Liendo}}, \bibinfo {author} {\bibfnamefont {C.}~\bibnamefont {Meneghelli}}, \ and\ \bibinfo {author} {\bibfnamefont {V.}~\bibnamefont {Mitev}},\ }\href {\doibase 10.1007/JHEP10(2018)077} {\bibfield  {journal} {\bibinfo  {journal} {JHEP}\ }\textbf {\bibinfo {volume} {10}},\ \bibinfo {pages} {077} (\bibinfo {year} {2018})},\ \Eprint {http://arxiv.org/abs/1806.01862} {arXiv:1806.01862 [hep-th]} \BibitemShut {NoStop}%
\bibitem [{\citenamefont {Green}\ \emph {et~al.}(2010)\citenamefont {Green}, \citenamefont {Komargodski}, \citenamefont {Seiberg}, \citenamefont {Tachikawa},\ and\ \citenamefont {Wecht}}]{Green:2010da}%
  \BibitemOpen
  \bibfield  {author} {\bibinfo {author} {\bibfnamefont {D.}~\bibnamefont {Green}}, \bibinfo {author} {\bibfnamefont {Z.}~\bibnamefont {Komargodski}}, \bibinfo {author} {\bibfnamefont {N.}~\bibnamefont {Seiberg}}, \bibinfo {author} {\bibfnamefont {Y.}~\bibnamefont {Tachikawa}}, \ and\ \bibinfo {author} {\bibfnamefont {B.}~\bibnamefont {Wecht}},\ }\href {\doibase 10.1007/JHEP06(2010)106} {\bibfield  {journal} {\bibinfo  {journal} {JHEP}\ }\textbf {\bibinfo {volume} {06}},\ \bibinfo {pages} {106} (\bibinfo {year} {2010})},\ \Eprint {http://arxiv.org/abs/1005.3546} {arXiv:1005.3546 [hep-th]} \BibitemShut {NoStop}%
\bibitem [{\citenamefont {Anselmi}(1999)}]{Anselmi:1998ms}%
  \BibitemOpen
  \bibfield  {author} {\bibinfo {author} {\bibfnamefont {D.}~\bibnamefont {Anselmi}},\ }\href {\doibase 10.1016/S0550-3213(98)00848-7} {\bibfield  {journal} {\bibinfo  {journal} {Nucl. Phys. B}\ }\textbf {\bibinfo {volume} {541}},\ \bibinfo {pages} {369} (\bibinfo {year} {1999})},\ \Eprint {http://arxiv.org/abs/hep-th/9809192} {arXiv:hep-th/9809192} \BibitemShut {NoStop}%
\bibitem [{\citenamefont {Belitsky}\ \emph {et~al.}(2008)\citenamefont {Belitsky}, \citenamefont {Henn}, \citenamefont {Jarczak}, \citenamefont {Mueller},\ and\ \citenamefont {Sokatchev}}]{Belitsky:2007jp}%
  \BibitemOpen
  \bibfield  {author} {\bibinfo {author} {\bibfnamefont {A.~V.}\ \bibnamefont {Belitsky}}, \bibinfo {author} {\bibfnamefont {J.}~\bibnamefont {Henn}}, \bibinfo {author} {\bibfnamefont {C.}~\bibnamefont {Jarczak}}, \bibinfo {author} {\bibfnamefont {D.}~\bibnamefont {Mueller}}, \ and\ \bibinfo {author} {\bibfnamefont {E.}~\bibnamefont {Sokatchev}},\ }\href {\doibase 10.1103/PhysRevD.77.045029} {\bibfield  {journal} {\bibinfo  {journal} {Phys. Rev. D}\ }\textbf {\bibinfo {volume} {77}},\ \bibinfo {pages} {045029} (\bibinfo {year} {2008})},\ \Eprint {http://arxiv.org/abs/0707.2936} {arXiv:0707.2936 [hep-th]} \BibitemShut {NoStop}%
\bibitem [{\citenamefont {Rychkov}\ and\ \citenamefont {Tan}(2015)}]{Rychkov:2015naa}%
  \BibitemOpen
  \bibfield  {author} {\bibinfo {author} {\bibfnamefont {S.}~\bibnamefont {Rychkov}}\ and\ \bibinfo {author} {\bibfnamefont {Z.~M.}\ \bibnamefont {Tan}},\ }\href {\doibase 10.1088/1751-8113/48/29/29FT01} {\bibfield  {journal} {\bibinfo  {journal} {J. Phys. A}\ }\textbf {\bibinfo {volume} {48}},\ \bibinfo {pages} {29FT01} (\bibinfo {year} {2015})},\ \Eprint {http://arxiv.org/abs/1505.00963} {arXiv:1505.00963 [hep-th]} \BibitemShut {NoStop}%
\bibitem [{\citenamefont {Alday}\ and\ \citenamefont {Maldacena}(2007)}]{Alday:2007he}%
  \BibitemOpen
  \bibfield  {author} {\bibinfo {author} {\bibfnamefont {L.~F.}\ \bibnamefont {Alday}}\ and\ \bibinfo {author} {\bibfnamefont {J.}~\bibnamefont {Maldacena}},\ }\href {\doibase 10.1088/1126-6708/2007/11/068} {\bibfield  {journal} {\bibinfo  {journal} {JHEP}\ }\textbf {\bibinfo {volume} {11}},\ \bibinfo {pages} {068} (\bibinfo {year} {2007})},\ \Eprint {http://arxiv.org/abs/0710.1060} {arXiv:0710.1060 [hep-th]} \BibitemShut {NoStop}%
\bibitem [{\citenamefont {Ferrero}\ and\ \citenamefont {Meneghelli}(2024)}]{Ferrero:2023znz}%
  \BibitemOpen
  \bibfield  {author} {\bibinfo {author} {\bibfnamefont {P.}~\bibnamefont {Ferrero}}\ and\ \bibinfo {author} {\bibfnamefont {C.}~\bibnamefont {Meneghelli}},\ }\href {\doibase 10.1007/JHEP05(2024)090} {\bibfield  {journal} {\bibinfo  {journal} {JHEP}\ }\textbf {\bibinfo {volume} {05}},\ \bibinfo {pages} {090} (\bibinfo {year} {2024})},\ \Eprint {http://arxiv.org/abs/2312.12550} {arXiv:2312.12550 [hep-th]} \BibitemShut {NoStop}%
\bibitem [{\citenamefont {Cooke}\ \emph {et~al.}(2017)\citenamefont {Cooke}, \citenamefont {Dekel},\ and\ \citenamefont {Drukker}}]{Cooke:2017qgm}%
  \BibitemOpen
  \bibfield  {author} {\bibinfo {author} {\bibfnamefont {M.}~\bibnamefont {Cooke}}, \bibinfo {author} {\bibfnamefont {A.}~\bibnamefont {Dekel}}, \ and\ \bibinfo {author} {\bibfnamefont {N.}~\bibnamefont {Drukker}},\ }\href {\doibase 10.1088/1751-8121/aa7db4} {\bibfield  {journal} {\bibinfo  {journal} {J. Phys. A}\ }\textbf {\bibinfo {volume} {50}},\ \bibinfo {pages} {335401} (\bibinfo {year} {2017})},\ \Eprint {http://arxiv.org/abs/1703.03812} {arXiv:1703.03812 [hep-th]} \BibitemShut {NoStop}%
\bibitem [{\citenamefont {Komargodski}\ and\ \citenamefont {Simmons-Duffin}(2017)}]{Komargodski:2016auf}%
  \BibitemOpen
  \bibfield  {author} {\bibinfo {author} {\bibfnamefont {Z.}~\bibnamefont {Komargodski}}\ and\ \bibinfo {author} {\bibfnamefont {D.}~\bibnamefont {Simmons-Duffin}},\ }\href {\doibase 10.1088/1751-8121/aa6087} {\bibfield  {journal} {\bibinfo  {journal} {J. Phys. A}\ }\textbf {\bibinfo {volume} {50}},\ \bibinfo {pages} {154001} (\bibinfo {year} {2017})},\ \Eprint {http://arxiv.org/abs/1603.04444} {arXiv:1603.04444 [hep-th]} \BibitemShut {NoStop}%
\bibitem [{\citenamefont {Behan}(2018)}]{Behan:2017mwi}%
  \BibitemOpen
  \bibfield  {author} {\bibinfo {author} {\bibfnamefont {C.}~\bibnamefont {Behan}},\ }\href {\doibase 10.1007/JHEP03(2018)127} {\bibfield  {journal} {\bibinfo  {journal} {JHEP}\ }\textbf {\bibinfo {volume} {03}},\ \bibinfo {pages} {127} (\bibinfo {year} {2018})},\ \Eprint {http://arxiv.org/abs/1709.03967} {arXiv:1709.03967 [hep-th]} \BibitemShut {NoStop}%
\bibitem [{\citenamefont {Polchinski}\ and\ \citenamefont {Sully}(2011)}]{Polchinski:2011im}%
  \BibitemOpen
  \bibfield  {author} {\bibinfo {author} {\bibfnamefont {J.}~\bibnamefont {Polchinski}}\ and\ \bibinfo {author} {\bibfnamefont {J.}~\bibnamefont {Sully}},\ }\href {\doibase 10.1007/JHEP10(2011)059} {\bibfield  {journal} {\bibinfo  {journal} {JHEP}\ }\textbf {\bibinfo {volume} {10}},\ \bibinfo {pages} {059} (\bibinfo {year} {2011})},\ \Eprint {http://arxiv.org/abs/1104.5077} {arXiv:1104.5077 [hep-th]} \BibitemShut {NoStop}%
\bibitem [{\citenamefont {Beccaria}\ \emph {et~al.}(2022)\citenamefont {Beccaria}, \citenamefont {Giombi},\ and\ \citenamefont {Tseytlin}}]{Beccaria:2022bcr}%
  \BibitemOpen
  \bibfield  {author} {\bibinfo {author} {\bibfnamefont {M.}~\bibnamefont {Beccaria}}, \bibinfo {author} {\bibfnamefont {S.}~\bibnamefont {Giombi}}, \ and\ \bibinfo {author} {\bibfnamefont {A.~A.}\ \bibnamefont {Tseytlin}},\ }\href {\doibase 10.1088/1751-8121/ac7018} {\bibfield  {journal} {\bibinfo  {journal} {J. Phys. A}\ }\textbf {\bibinfo {volume} {55}},\ \bibinfo {pages} {255401} (\bibinfo {year} {2022})},\ \Eprint {http://arxiv.org/abs/2202.00028} {arXiv:2202.00028 [hep-th]} \BibitemShut {NoStop}%
\end{thebibliography}%


\begin{thebibliography}{0}%
\makeatletter
\providecommand \@ifxundefined [1]{%
 \@ifx{#1\undefined}
}%
\providecommand \@ifnum [1]{%
 \ifnum #1\expandafter \@firstoftwo
 \else \expandafter \@secondoftwo
 \fi
}%
\providecommand \@ifx [1]{%
 \ifx #1\expandafter \@firstoftwo
 \else \expandafter \@secondoftwo
 \fi
}%
\providecommand \natexlab [1]{#1}%
\providecommand \enquote  [1]{``#1''}%
\providecommand \bibnamefont  [1]{#1}%
\providecommand \bibfnamefont [1]{#1}%
\providecommand \citenamefont [1]{#1}%
\providecommand \href@noop [0]{\@secondoftwo}%
\providecommand \href [0]{\begingroup \@sanitize@url \@href}%
\providecommand \@href[1]{\@@startlink{#1}\@@href}%
\providecommand \@@href[1]{\endgroup#1\@@endlink}%
\providecommand \@sanitize@url [0]{\catcode `\\12\catcode `\$12\catcode `\&12\catcode `\#12\catcode `\^12\catcode `\_12\catcode `\%12\relax}%
\providecommand \@@startlink[1]{}%
\providecommand \@@endlink[0]{}%
\providecommand \url  [0]{\begingroup\@sanitize@url \@url }%
\providecommand \@url [1]{\endgroup\@href {#1}{\urlprefix }}%
\providecommand \urlprefix  [0]{URL }%
\providecommand \Eprint [0]{\href }%
\providecommand \doibase [0]{http://dx.doi.org/}%
\providecommand \selectlanguage [0]{\@gobble}%
\providecommand \bibinfo  [0]{\@secondoftwo}%
\providecommand \bibfield  [0]{\@secondoftwo}%
\providecommand \translation [1]{[#1]}%
\providecommand \BibitemOpen [0]{}%
\providecommand \bibitemStop [0]{}%
\providecommand \bibitemNoStop [0]{.\EOS\space}%
\providecommand \EOS [0]{\spacefactor3000\relax}%
\providecommand \BibitemShut  [1]{\csname bibitem#1\endcsname}%
\let\auto@bib@innerbib\@empty
%</preamble>
\end{thebibliography}%

\end{document}